\documentclass[12pt]{article} 
\usepackage[a4paper]{typearea}
\typearea{0}
\usepackage{helvet}
\usepackage{graphicx}
\usepackage{amsmath}
\usepackage{times}
\usepackage {graphics,pstricks}
\usepackage[T1]{fontenc}
\usepackage[latin1]{inputenc}
\usepackage {pst-all}
\usepackage {multido}
\usepackage{color}
\usepackage{natbib}

\begin{document}
\def\sini{\frac{\mid \Delta\varepsilon\rangle}{\Delta\varepsilon}}
\def\siniii{\frac{\mid 3\Delta\varepsilon\rangle}{3\Delta\varepsilon}}
\def\sinv{\frac{\mid 5\Delta\varepsilon\rangle}{5\Delta\varepsilon}}
\def\sint{{\rm sin (\Delta\varepsilon\it t/\hbar)}}
\def\sinttt{\frac{\rm sin (3\Delta\varepsilon\it t/\hbar)}{3}}
\def\sinvt{\frac{\rm sin (5\Delta\varepsilon\it t/\hbar)}{5}}
\def\cost{{\rm cos (\Delta\varepsilon\it t/\hbar)}}
\def\costtt{{\rm cos (3\Delta\varepsilon\it t/\hbar)}}
\def\cosvt{{\rm cos (5\Delta\varepsilon\it t/\hbar)}}
\def\wkgm{\mid w_{\kappa g}-\rangle}
\def\DV{$\Delta V$}
\def\sqwv{\sqrt{NW^2+(\Delta V)^2}}

\newcommand{\trdim}{3 dimensional }
\newcommand{\fdim}{4 dimensional }
\newcommand{\fD}{4 dimensions }
\newcommand{\edim}{11 dimensional }
\newcommand{\SE}{Schr\"odinger equation }
\newcommand{\ST}{Schr\"odinger theory }
\newcommand{\CQM}{Copenhagen quantum mechanics }
\newcommand{\cs}{configuration space }
\newcommand{\wfg}{weak field gravitation }
\newcommand{\qft}{quantum field theory }
\newcommand{\vl}{velocity of light }
\newcommand{\tsl}{telegraph signal like }

\begin{center}
\Large{Copenhagen Quantum Mechanics Emerges from a Deterministic Schr\"odinger 
Theory in 11 Dimensional Spacetime Including Weak Field Gravitation}\\
\vspace{.5cm}
{\large G. Doyen$^1$ D. Drakova$^2$}  \\

\vspace{.5cm}
{\small
$^1$ Ludwig-Maximilians Universit\"at, M\"unchen, Germany \\
URL: gerold@gerold-doyen.de\\
$^2$ Faculty of Chemistry, University of Sofia, Bulgaria\\
URL: nhdd@chem.uni-sofia.bg} \\
\end{center}

\begin{abstract}
We construct a world model consisting of matter fields living in \fdim
spacetime and a gravitational field living in \edim spacetime.
The seven hidden dimensions are compactified within a radius
estimated by reproducing the particle - wave characteristic of
diffraction experiments.
In the presence of matter fields
the gravitational field develops localized modes with elementary excitations
called gravonons which are induced by the sources
(massive particles).
The final world model treated here
contains only gravonons,
gravitons and a scalar matter field. The gravonons are localized in the
environment of the massive
particles which generate them. The solution of the \SE for the world
model yields
matter fields which are localized in the 4 dimensional subspace. The
localization has the following properties:
(i) There is a chooser mechanism for the selection of the localization
site. (ii) The chooser
selects one site on the basis of minor energy and gravonon differences
between the sites, which at present
cannot be controlled experimentally and therefore let the choice appear
statistical. (iii) The changes from one localization site to a
neighbouring one take place
in a telegraph-signal like manner. (iv) The times at which telegraph
like jumps occur
dependent on subtleties of the gravonon structure which at present cannot be
controlled experimentally and therefore let the telegraph-like jumps
appear statistical. (v) The fact that the dynamical law acts in the \cs
of fields living in \edim spacetime lets the events observed in \fdim
spacetime appear non-local. 
In this way the phenomenology of \CQM is obtained without the need
of introducing the process of collapse and a statistical (probabilistic)
interpretation of the wave
function. Operators defining observables need not be introduced. The
known commutation
relations for observables in \fD can be derived but they are no prerequisite
for interpreting experiments. All experimental findings are explained in
a deterministic
way as a consequence of the time development of the wave function in \cs
according to Schr\"odinger's equation
without the need of introducing a statistical interpretation. \\

\vspace{.5cm}
\end{abstract}

\section{Introduction}
\label{intro}
The problem of whether quantum mechanics, including the postulated 
concept of random collapses, provides a real representation of the world
or whether it implies just a probabilistic interpretation of experimental data
is still discussed controversially.   
Whereas Penrose   
suggests that collapse is a dynamics arising in a natural way from gravity 
\cite{penrose1}-\cite{penrose6},
Omnes claims that decoherence solves the problem and, hence, 
objectification in quantum mechanics is a non-existent problem \cite{omnes}. 
Kiefer and Joos maintain that the
dynamical collapse models invoking gravity may be simply drowned by   
environmental decoherence and hence are to be considered as ''excess baggage'' \cite{kiefer1}-\cite{kieferjoos}. 
As far as locality is concerned,  
d'Espagnat contends that decoherence theory does not reconcile
physics with the objective existance of phenomena (strong realism) \cite{despagnat1,despagnat2}.
D\"urr and Teufel maintain that decoherence does    
not create the facts of our world, but rather produces a sequence   
of fapp-redundancies, which physically increase or stabilize decoherence. They believe
that physical theory should describe the behaviour of real objects, located
in physical space \cite{duerrbook}.
In the present paper we report a theory which is in the spirit
of quantum realism.\\

\noindent As is well known and well accepted, the \SE evolves a state, which qualifies
as representing real world, into a state which is a superposition of real
world states and as such does not qualify as a real world state.
To handle this problem von Neumann \cite{vonNeumann} introduced
a second dynamics which is not unitary and serves as reduction
on real world states. This second dynamics is referred to as collapse
or state vector reduction and cannot presently be described by a
mathematical theory. We refer to this state of 
quantum theory as Copenhagen Quantum Mechanics (CQM). \\

\noindent
In recent years there have been attempts to construct theories where
\CQM and/or \qft emerge in the low energy (infrared) limit of a
(higher-dimensional) theory which is capable of combining quantum
mechanics and general relativity. Superstring theory and/or M-theory
\cite{string} are the most well-known examples.
According to a different theory developed by 't Hooft the fundamental
laws valid at the Planck scale are
derived from Newton's law, but are different. In this theory \CQM
emerges as the low energy limit
due to a chaotic (dissipative) dynamics (loss of information) with a set
of attractors operating during the development of the system
from the Planck scale to the microscopic (atomic) scale. Classical
variables can assume
any arbitrary value out of a continuum while quantum states are discrete
\cite{thooft1}-\cite{thooft3}.
Inspired by 't Hooft's idea,
Blasone et al. \cite{blasone} show that the
combined evolution of two classical, dissipative harmonic oscillators
results
in a linear harmonic quantum oscillator with a zero point energy.\\

\noindent
Adler proposes that \CQM is an emergent phenomenon arising from a deeper
level
of dynamics via an extension of classical dynamics to non-commuting
matrix variables, with cyclic permutation inside a trace used as the
basic calculational tool.
\CQM is then shown to emerge as the statistical thermodynamics of this
underlying
theory, with the canonical commutation-anticommutation relations derived
from a
generalized equipartition theorem. State vector reduction and probabilistic
interpretation of \CQM are then argued to follow from Brownian motion
like corrections to the \SE
\cite{adler,adler1}. Smolin considers classical matrix models with an explicit
stochastic noise giving
rise to quantum behaviour \cite{smolin}.\\

\noindent
Bir\'o et al. \cite{biro} demonstrate that a classical (Euclidean) field
theory living in five
dimensions can behave in the infrared limit like a quantum system, if it
is only observed
in four dimensions.
Hadley argues that the fundamental logic of \CQM results from general
relativity.
In his theory elementary particles are not separate objects living in
\fdim spacetime, but
rather constituents (local deformations) of spacetime. General
relativity contains
past and future in a deterministic manner. The statistical character of
\CQM is then a
consequence of the missing knowledge of the future \cite{hadley}.\\

\noindent 
A different approach is to modify the \SE in such a way that the collapse
dynamics is included in its solution. These are the so called collapse models
\cite{GRW,Pearlecoll}. For a detailed review see ref. \cite{Pearlecoll}.\\

\noindent It is often stated that \CQM explains all known phenomena and
experimental findings
and that experimental hints to a different physics are not available
\cite{haag}.
This is clearly not true.
With the advent of experimental techniques, which
provide data with time resolution and spatial resolution on the atomic
scale
of the dynamics of chemical and physical processes on solid surfaces
in a non-destructive way, \CQM is challenged.
Within \CQM collapse of the total wave function might be postulated as the
process leading to the localization of adsorbed atoms and molecules on
solid surfaces,
as they are imaged in experiments with the low-temperature
scanning tunnelling microscope \cite{eiglerXe,lauhonho}. The two-dimensional
periodicity of the solid surfaces means delocalization of the wave
function in two dimensional Bloch
waves resulting
from the time development of the wave packets according to
Schr\"odinger's equation.
It would imply that an initially localized adparticle should after some time
be observed anywhere on equivalent sites on the surface with equal probability.
This is, however, never observed.
An extension of \CQM by including the system and the environment is often
considered under the heading of decoherence theory.
According to this theory localization via interaction, ''permanent
measurement''
by environmental particles
(phonons, tomonagons, photons, plasmons, etc.) is suggested. But it
cannot be the explanation
either.
This is so because at low temperatures the available environmental
excitations have very long
wavelengths compared to the size of the adsorbed particles.
Furthermore, when adsorbed particles jump between two adsorption sites,
the movement occurs as telegraph-signal like jumps and not as the smooth
Rabi oscillations
predicted by Schr\"odinger's equation \cite{meyer}.
The diffusion rates in the quantum diffusion regime of the same
adsorbate on similar metal surfaces,
measured in different experiments,
differ by several orders of magnitude \cite{lauhonho,gomer}. Copenhagen quantum mechanics,
with the coupled dynamics of the adsorbate motion and the substrate
phonons taken into account,
provides results, which drastically differ by nearly nine orders of
magnitude from the experimental
diffusion rates \cite{wahn1997}.
The desorption
reaction of CO, induced by tunnelling electrons in the
scanning tunnelling microscope \cite{COdesorption}, can be explained
only by assuming that the tunnelling electron is transiently localized in
the close vicinity of the adsorbate \cite{doydrakCOdesorp}.
The survival of metastable atoms in interaction with adsorbate covered solid
surfaces \cite{ertl} presents a further problem, which cannot be
resolved within Copenhagen quantum mechanics.
The survival probability of metastable atoms He$^*$ on
a metal surface, covered with CO, is attenuated compared to the
deexcitation probability in
a scattering experiment between a beam of He$^*$ and CO in the gas phase,
which is unexpected and not explained within \CQM upto now.
In ref. \cite{doyHe-COscattering} an understanding has been suggested,
based on the different dimensionality of
the scattering continua in the gas phase and on the solid surface.
This has been termed the dimensionality effect \cite{doyphyrevb} and is
in the spirit of the theory developed here.\\

\noindent Recently, a programmable quantum annealing machine has been
built which
uses quantum effects to minimize a cost function \cite{aepli}.
The question has been raised whether the behaviour of this D-wave
computer \cite{dwavesys}
can be described by classical statistical mechanics, an approach called
simulated annealing, or
whether \CQM has to be invoked in order to explain the operation
of D-wave. It has been established experimentally by Chiorescu et al.
\cite{chiorescu}
that a single qubit loses coherence within nanoseconds,
i.e. within a few nanoseconds the behaviour of a single qubit
can no longer be described by Copenhagen quantum mechanics.
In Ramsey interferometry experiment a $\frac{\pi}{2}$ microwave pulse
initializes Rabi oscillations between the macroscopic supercurrent 
in the flux qubit in two reverse directions,  
which are interpreted as
oscillating probability for occupation of one current state. The envelope
of the oscillations of this probability decays, providing a
''coherence time'' of one current state of
the order of 20 nanoseconds.
These observations are interpreted to demonstrate that within
nanoseconds the behaviour of a single qubit
can no longer be described by Copenhagen quantum mechanics.\\

\noindent On the other hand, Johnson et al.
show that in a macroscopic system of an eight-qubit chain the quantum
annealing of
the system at low temperature results in quantum tunnelling in the
global energy minimum,
with a temperature independent freezing time \cite{johnson}.
Furthermore the experiment of Dickson et al. \cite{dwave} using 16 qubits of
a superconducting quantum processor and
quantum annealing with the help of applied transverse magnetic fields
shows that, even with annealing time eight orders of magnitude
longer than the time, determined for the decay of the oscillations
described above,
the system behaves as predicted by Copenhagen quantum mechanics.
Classical models do not reproduce
the experimental results. \\

\noindent 
We applied the theory presented in this contribution successfully
to explain
{\em why does the D-wave computer
operate as a quantum computer over minutes, despite that the
effects of the thermal environment are on a timescale of nanoseconds}.
An understanding can be provided in the framework of the present theory \cite{unpublished}.\\

\noindent It appears that the localization of quantum particles via
entanglement
to environmental continua of high density of states
is a necessary condition for various reactions on solid surfaces.
Particle localization in a
collapse process, as suggested in \CQM is, however, not the result
of the solution of Schr\"odinger's equation or of any well defined dynamics.
The examples from the previous paragraph illustrate that on a nanoscale
level on solid surfaces there are
dynamical processes which cannot be accounted for by Copenhagen quantum mechanics.
In all cases mentioned, entanglement with
an environment obviously plays a role. However, the environmental
excitations usually discussed in the context of these low temperature
experiments,
for instance phonons, tomonagons, plasmons etc., are not capable
to localize atoms or electrons on a solid surface because of their long
wavelength. A new explanation is needed.\\

\noindent
We develop a theory
where \CQM emerges from a quantum field theory, which treats
gravitation in eleven spacetime dimensions in the weak field limit.
Entanglement to soft modes in the hidden spatial dimensions
leads to particle localization,
which is a necessary condition for the processes on the atomic scale,
listed in the previous paragraph.
Changes of localization sites occur as a telegraph like dynamics
with entanglement to low frequency modes
and in the limit of vanishing interaction strength.
Non-locality in four spacetime dimensions is implicit in the theory. It also is capable of
describing and explaining the
experiments on solid surfaces mentioned beforehand, which was found
impossible within Copenhagen quantum mechanics.\\

\noindent There exist many attempts to account for the effects of gravity on a quantum system 
by modifying Schr\"odinger's equation including stochastic   
attractive Newtonian gravitational fields in 4 dimensional spacetime \cite{diosi3}-\cite{hughston}
or introducing a nonlinear ''Schr\"odinger-Newton'' equation \cite{penrose1,diosi3}. 
The interpretation is based      
on the tendency of the off-diagonal elements of the reduced density matrix  
of a quantum system in interaction with the external gravitational field to
attenuate, and is assumed to imply transition to classical and random behaviour. 
Instead, gravitation has been suggested as the
origin of the noise field in collapse models \cite{Pearlecoll}. Penrose's intensive attempts
also suggest gravitation as the origin of collapse \cite{penrose1}-\cite{penrose6}. 
With the purpose to describe localization and transition of the quantum mechanical
superposition of states to classical states,
even the decay mecahanism
is often postulated \cite{penrose1,diosi3}.
Within the framework of continuous spontaneous reduction models   
\cite{GRW} Pearle and Squires interpret the classical scalar field, which causes   
collapse, as the gravitational curvature scalar, however,
gravity is treated semiclassically \cite{pearle}.   
In the work of Anastopoulos the gravitational field
is considered as leading to classical behaviour,  
however, not for microscopic particles \cite{anast}. Spacetime fluctuations due to 
stochastic backgrounds of gravitational waves  
\cite{lamine} or metric fluctuations \cite{breuer} have also been suggested  
to lead to unavoidable transition of quantum particles to classical behaviour.\\ 

\noindent
The aspects of \CQM are shown in this paper to emerge from
entanglement due to an extremely localized and weak interaction
between matter fields and a massless boson field of high mode density.
The exact form of the Lagrangian for the interacting fields will be
developed in sections 3 to 5. In section 2 we investigate the structure
of a theory based on an extremely local and weak interaction with a high
density continuum in \edim spacetime. This includes compactification of
the hidden
dimensions (section \ref{hidden}), determination of the density of states
(section \ref{gravdos})
and the special kind of solution not obtainable in perturbation theory
(sections \ref{charsolution} and \ref{telegraph}).
The properties of the solution are exemplified for the case of diffraction
from nano lattices. In sections \ref{ontologyEQM} and \ref{comparisonCQMEQM}
the ontology of our Emerging Quantum Mechanics (EQM) is set in relation 
with that of Copenhagen and Bohmian quantum mechanics. 

\section{The structure of a deterministic \ST in \edim spacetime
including \wfg }    
As mentioned in the introduction a possible theory for the unification
of general
relativity and quantum mechanics is superstring theory and/or M-theory
\cite{string}.
The equations of superstring theory are mathematically consistent only,
if the strings
move in 10 dimensional space. According to M-theory spacetime can
besides
one-dimensional strings also contain branes of various dimensions.
Strings can be
restricted to subspaces of spacetime, if they are bound to branes. The
physical laws
depend on the geometry and dimensions of the branes and the hidden
dimensions.

\subsection{Hidden dimensions}
First we briefly summarize the way we think about this subject.
A single hidden dimension can only have the shape of a circle. More
extra dimensions
can have a variety of shapes (called topologies) such as spheres, tori,
connected tori,
etc. differing in length and diameter. This results in a gigantic number
of possible
geometries. In addition there are parameters describing the positions of
the branes
and the fluxes around the tori. Each configuration of branes and hidden
dimensions
has a different energy. If there are no fields in \fdim spacetime this corresponds
to the
vacuum energy. For a given three dimensional brane the geometry of the hidden
dimensions is determined by
minimizing
this vacuum energy. The multitude of possible geometries results in the
so called
landscape of string theory. Our universe has to correspond to a minimum
on the landscape
with a relatively small vacuum energy, the geometry of branes and hidden
dimensions
corresponding to our universe is, however, unknown. For the purpose of this
investigation we assume the simplest possible case, namely a flat 3-dimensional
brane and a seven
dimensional sphere representing the hidden dimensions. All known fields
except
for the gravitational field are assumed to live only on the three dimensional brane.
Except for
the existence of such a three dimensional brane and the spherically compactified
hidden dimensions,
no other properties of superstring theory and/or M-theory are needed.

\subsubsection{Compactification of the hidden dimensions}\label{hidden}
The basic idea of how \CQM emerges in our theory is that the
gravitational interaction provides {\em a chooser} that guides a diffracted
matter field to a particular site on the screen. Assume the simple
case that a point like source emitting the matter field is exactly
in the center of a three dimesional sphere of radius 1 meter. The area of the sphere is
then $4\pi$ m$^2 \approx 16\pi 10^{20}$ bohr$^2 \approx 5\cdot
10^{21}$ bohr$^2$.
Assuming an area of the lattice unit cell of the screen of 25 bohr$^2$,
there will be $2\cdot 10^{20}$ sites on the screen where the matter field
could localize ("collapse"). The idea is that the sites are fully equivalent
but vary slightly in energy and geometry. A realistic estimate for this
is a variation in energy of $\approx 10^{-3}$ eV and a variation in geometry
of $\approx 10^{-1}$ bohr.\\

\noindent
The average spacing between two adjacent energy levels is
$10^{-3}$ eV$/10^{20}$ sites$=10^{-23}$ eV.
If the only criterion is energy then the chooser has to distinguish energy
levels on a scale of $10^{-23}$ eV in order to select a single site.
Analogously the average deviation between two similar geometries will
be $10^{-1}$ bohr$/10^{20}$ sites$=10^{-21}$ bohr. If the only criterion is
geometry
then the chooser has to distinguish geometry on a scale of
$10^{-21}$ bohr in order
to select a single site. If we assume that the chooser is sensitive to
geometric
differences of $10^{-5}$ bohr then $10^{16}$ sites would be available for
a choice.
This means then that the chooser should distinguish on an energy scale of
$10^{-3}$ eV$/10^{16}$ sites$=10^{-19}$ eV in order to select a single site.\\

\noindent
The consequence is that the interaction energy between a quantum of the
matter
field and a site on the screen should be of this order of magnitude. A
larger interaction
energy cannot lead to site selection. A significantly smaller
interaction energy
would mean that either very many quanta of the matter field have to be
emitted from
the source, before a "collapse" can be registered, or that it would take
months or years
before a quantum can be registered on the screen. Both cases violate
strongly
the experimental findings.\\

\noindent
The required interaction strength is much weaker than any kind of
electrodynamical, chemical or
van der Waals interaction. This means that coupling to phonons or
electron - hole pairs, i.e. 
"measurement by the environment", is not capable of
describing
the "collapse" or measurement process which is at the heart of the so called
wave - particle duality. On the other hand the required interaction
strength is much
stronger than gravitational interaction in \fdim spacetime. Consider two
protons at a distance
of r=6 bohr and mass $M=10^4$ a.u. Their gravitational interaction energy is
$GM^2/r\approx
10^{-33}$ eV,
where we inserted a value of $G=10^{-40}$ atomic units for Newton's
gravitational
constant. This is 14 orders of magnitude smaller than required. The only
possibility for
obtaining the required order of magnitude for the interaction energy
appears to be
a gravitational field in \edim spacetime with compactified hidden
dimensions.\\

\noindent
The gravitational potential provided
by a pointlike mass $M$ in 4 dimensional spacetime is:
{
\begin{equation}\label{classgrav}
V_{grav}^{(4)}(r) = -\frac{GM}{r}
\end{equation}}
In 11 dimensional spacetime (10 space
dimensions) the gravitational potential is:\index{hidden dimensions}
\begin{equation}\label{hiddengrav}
V_{grav}^{(11)}(r) = -\frac{G^{(11)}M}{\pi^7r^8}
\end{equation}
The gravitational law eq. (\ref{hiddengrav}) cannot be valid
for large separations $r$ as this would violate the experimentally
verified classical law eq. (\ref{classgrav}). Therefore the hidden
dimensions
are rolled up (compactified)\index{compactified dimensions}
to a diameter $2a$ so that at large distances the
separation in the hidden dimensions never exceeds $2a$.
Equating the classical and the 11 dimensional gravitational
law at large distances, we obtain
\begin{equation}\label{g11}
-\frac{GM}{r} =    -\frac{G^{(11)}M}{(2a\pi )^7r}  \rightarrow
G^{(11)}=(2a\pi )^7G
\end{equation}
This choice makes the two laws eq. (\ref{classgrav}) and
eq. (\ref{hiddengrav})
agree at separations larger than $2a$. Inserting eq. (\ref{g11})
in eq. (\ref{hiddengrav}) yields at $r=1$ bohr:
\begin{equation}
V_{grav}^{(11)}(r=1) =- GM(2a)^7
\end{equation}
This means that at $r=1$ bohr the gravitational interaction
is $(2a)^7$ times stronger than predicted by the
classical law eq. (\ref{classgrav}).\\

\noindent
The dependence of $G^{(11)}$ on the compactification radius is then:

\begin{tabular}{lc}
a       &$G^{(11)}/\pi^7$\\
$10^4$&$10^{-10}$\\
$10^3$&$10^{-17}$\\
$10^2$&$10^{-24}$\\
$10$&$10^{-31}$\\
\end{tabular}

\noindent
From the selection criterion described above
we require $\frac{G^{(11)}M}{\pi^7r^8}\approx 10^{-20}$ atomic units at
$r=6$ bohr and $M=10^4$,
hence $\frac{G^{(11)}M}{\pi^76^8}\approx
\frac{G^{(11)}10^4}{\pi^710^6}\approx
\frac{G^{(11)}}{\pi^7}*10^{-2}\approx 10^{-20}$
which implies $\frac{G^{(11)}}{\pi^7}\approx 10^{-18}$.
For our chooser we would therefore need a compactification radius of
roughly $10^3$ bohr.
This is enormously larger than the Planck length. A chooser needs large
extra dimensions. In the theory of Arkani-Hamed et al.
it is also claimed that the hidden dimensions need not be compactifed on
the Planck scale, they can be as large as a fraction of a millimeter
or even infinite \cite{add4}.

\subsubsection{Graviton mode density} \label{gravdos}
The chooser can only work, if gravitons are available at an energy
separation
which is of the order of the gravitational interaction strength for any kind
of matter field quantum. This is only possible, if the density of
graviton modes
is sufficiently large.
The density of graviton modes has been calculated in ref. \cite{dice2010-more}.
Due to its importance the argument is repeated here. The calculation
runs parallel to the evaluation of the mode density
in the electromagnetic case. We expand the gravitational radiation
field in the eigenmodes of resonators with fixed frequency, polarization,
and field distribution. Although the selection of modes is in
principle arbitrary, the most popular expansion uses the eigenmodes
of rectangular boxes with perfectly reflecting walls. The mirrors at
position 0 and $a$ impose the boundary conditions that the field
vanishes at those positions. The eigenmodes have field distributions that
vary with ${\rm sin}(n\pi x/a)$, where $n$ is the mode number, $x$ the spatial
variable and $a$ the separation of the two mirrors. The spatiotemporal
variation of the field is
\begin{equation}
F_n={\rm sin}(n\pi x/a){\rm e}^{i\omega_nt}
\end{equation}
The wave number $k=(n\pi /a)$
determines the angular frequency $\omega_n=\frac{n\pi {\rm c}}{a}$.
The one dimensional density of modes in $k$-space is
\begin{equation}\label{rhok}
\rho_k=\frac{a}{\pi }.
\end{equation}
Using the dispersion relation for gravitons $\epsilon_\kappa =\kappa {\rm c}$, 
with ${\rm c}$ the velocity of light, \\
$\kappa = \sqrt{k_1^2+k_2^2+k_3^2+k_4^2+...+k_d^2}$
and $d$ the dimension of gravitational $k$-space, we write for the graviton
mode density
\begin{equation}
\rho_{grav}(E)=\sum_\kappa \delta(E-\epsilon_\kappa ) =
\int {\rm d}^d\kappa \rho_\kappa^d\delta(E-\epsilon_\kappa ).
\end{equation}
Transforming to spherical coordinates $\eta$ one obtains \cite{zwiebach};
\begin{equation}
\rho_{grav}(E)=\int {\rm d}\eta\eta^{d-1}\delta(E-\epsilon_{\kappa})\rho_\kappa^d
\frac{\pi^{d/2}}{\Gamma (1+\frac{d}{2})}. 
\end{equation}
Specializing to 10 spatial dimensions and substituting
${\rm d}\eta ={\rm d}\epsilon /{\rm c}$ one obtains
\begin{eqnarray}
\rho_{grav}(E)&=&\frac{1}{{\rm c}}\frac{\pi^5}{\Gamma (1+5)}\rho_\kappa^d
\int {\rm d}\epsilon\frac{\epsilon^{d-1}}{{\rm c}^{d-1}}\delta (E-\epsilon)\\
&=& \frac{E^{9}}{{\rm c}^{10}}\frac{\pi^5}{5!}(\frac{L}{\pi })^3(\frac{a}{\pi
})^{7} 
\end{eqnarray}
where $L$ is the normalization length of the macroscopic three dimensional space
and $a$ is the normalization length of the hidden dimensions.\\

\noindent
The density of states for the two dimensional movement of
an adparticle of mass $M$ is $\rho_{2D}=\frac{2ML^2}{\pi}$
which is of the order of $10^{17}$ for a proton.
Therefore the relationship between the mode density                                         
of gravitons at energy $E$ and the two dimensional adparticle
motion is
\begin{equation}
\frac{\rho_{grav}}{\rho_{2D}}=
\frac{\frac{E^{9}}{{\rm c}^{10}}}{2M}\frac{\pi^5}{5!}
\frac{La^{7}}{\pi^{9}}\approx 10^{34}. 
\end{equation}                       
For the wave vector of the graviton we assumed $\kappa\approx
10$ bohr$^{-1}$,
which corresponds to a wave length of 1 bohr
($\lambda = 2\pi/\kappa\approx 1$ bohr), and 
$L\approx 10^7$ bohr. With the choice 
$a\approx 10^4$ bohr and $M=2000$  a.u.,
the graviton energy is of
order of $\epsilon_\kappa \approx 10^3$ Hartree.\\

\noindent
In conclusion the graviton continuum appears infinitely dense compared to
the continua describing any particle motion in our three dimensional space.

\subsection{Characteristics of the solution}\label{charsolution}
In \qft the standard way of solving a problem is via perturbation theory
(Feynman diagrams).
This approach does not work at all in our case. The reason is that the
solution we are looking for appears in the limit of very small (vanishing)
interaction strength
(cf. section \ref{hidden}). Expanding in terms of powers of a coupling
constant, which then tends to zero, yields zero in every order and for the
final
summation.
This problem is known in quantum field theory, where topological (soliton) solutions appear
for coupling constant tending to zero (cf. e.g. Zee \cite{zee} who
refers to this situation as "breaking the shackles of Feynman diagrams").
In our case the entanglement between the matter field
and gravitons becomes maximal in the (mathematical) limit, where the
interaction
strength $V_{grav}$ tends to zero while at the same time the graviton
density of states
$\rho_{grav}$ tends to $\infty$ so that $V_{grav}\rho_{grav}$ stays
non-zero and finite.\\

\noindent
The method we have to pursue is solving the \SE in configuration space.
In \qft the wave function becomes a wave functional, i.e. it depends
directly on the fields. Considering for illustrative purposes only
two boson fields, the scalar matter field
$\phi (t,x_1,x_2,x_3)$ and the gravitational field
$\zeta (t,x_1,x_2,...,x_{10})$, we have to solve
the \SE
\begin{eqnarray}
&& i\frac{\partial}{\partial t}\Psi \left(\phi (
t,x_1,x_2,x_3),\zeta
(t,x_1,x_2,...,x_{10})\right)  \nonumber \\
&& = {\rm H} \Psi \left(\phi ( t,x_1,x_2,x_3),\zeta
(t,x_1,x_2,...,x_{10})\right).
\end{eqnarray}
Reasonable solutions can be obtained
by adapting techniques, which have been developed in theoretical quantum
chemistry under the heading of "configuration interaction" (CI).\\

\noindent
Consider the set of fields $\{\phi^n \zeta^m\}$ forming a Hilbert space.
Expanding the fields in modes ${\phi_\alpha}$ and ${\zeta_\beta}$:
\begin{eqnarray}
\phi &=& \sum_{\alpha =1}^\infty C_\alpha\phi_\alpha\\
\zeta &=& \sum_{\beta =1}^\infty D_\beta\zeta_\beta
\end{eqnarray}
we write the wave functional in Dirac notation as
\begin{equation}\label{confsum}
\mid\Psi\rangle = \sum_{\{n_\alpha\},\{n_\beta\}}A_{\{n_\alpha\}\{n_\beta\}}
\otimes_{\alpha =1}^\infty\mid n_\alpha\rangle\otimes_{\beta
=1}^\infty\mid n_\beta\rangle
\end{equation}
with
\begin{eqnarray}
\mid n_\alpha\rangle&=&(\phi_\alpha(x))^{n_\alpha}\\
\mid n_\beta\rangle   &=&(\zeta_\beta(x))^{n_\beta}\\
\otimes_{\alpha =1}^\infty\mid n_\alpha\rangle &=&\mid
n_1\rangle\otimes\mid n_2\rangle\otimes\mid n_3\rangle\otimes\mid
n_4\rangle\otimes ...\\
\{n_\alpha\} &=& n_1n_2n_3n_4n_5 ...
\end{eqnarray}
The time dependence of the modes ${\phi_\alpha}$ and ${\zeta_\beta}$ is
just an arbitrary
phase factor which can be set equal to unity. The time dependence arising from the
relative phases is then contained in $A_{\{n_\alpha\}\{n_\beta\}}$.
A scalar product $\langle\{n_\alpha\}\{n_\beta\}\mid\{ n_\gamma\}\{
n_\delta\}\rangle $
is defined by\\
\begin{equation}
\langle\{n_\alpha\}\{n_\beta\}\mid\{ n_\gamma\}\{ n_\delta\}\rangle
= \prod_{\alpha ,\beta ,\gamma ,\delta =1}^\infty\delta_{n_\alpha
n_\gamma}\delta_{n_\beta n_\delta}
\end{equation}

\noindent
We refer to $\mid\{n_\alpha\}\{n_\beta\}\rangle$ as a (field) configuration.
Introducing creation and annihilation operators $a_\alpha^+$,
$a_\alpha$, $b_\beta^+$, $b_\beta$
in the usual way, one writes for the field configuration \\
\begin{equation}\label{prod}
\mid\{n_\alpha\}\{n_\beta\}\rangle = \prod_{\alpha
=1}^\infty\prod_{\beta =1}^\infty (a_\alpha^+)^{n_\alpha}
(b_\beta^+)^{n_\beta}\mid\{n_\alpha =0\}\{n_\beta =0\}\rangle.
\end{equation}

\noindent
The solution of the time dependent \SE is
\begin{equation}
\mid\Psi (t)\rangle = e^{-i{\rm H}t}\mid\Psi (t=0)\rangle
\end{equation}
where ${\rm H}$ is the Hamiltonian derived in sections 3 to 5. The time dependent
amplitude becomes
\begin{equation}
A_{\{n_\alpha\}\{n_\beta\}} (t) = \langle\{n_\alpha\}\{n_\beta\}\mid
e^{-i{\rm H}t}\mid\Psi (t=0)\rangle .
\end{equation}
The initial functional $\mid\Psi (t=0)\rangle$ can be expanded in the field
configurations:
\begin{equation}
\mid\Psi (t=0)\rangle = \sum_{\{ n_\gamma\}\{ n_\delta\}} B_{\{
n_\gamma\}\{ n_\delta\}}
\mid\{ n_\gamma\}\{ n_\delta\}\rangle .
\end{equation}
For the amplitude one obtains then
\begin{equation}
A_{\{n_\alpha\}\{n_\beta\}} (t) = \sum_{\{ n_\gamma\}\{ n_\delta\}}
B_{\{ n_\gamma\}\{ n_\delta\}}
\langle\{n_\alpha\}\{n_\beta\}\mid e^{-i{\rm H}t}\mid\{ n_\gamma\}\{
n_\delta\}\rangle .
\end{equation}
The configuration interaction matrix elements in the last equation
have to be evaluated using the hamiltonian ${\rm H}$ which consists of terms like
$h_{\alpha\beta}a_\alpha^+a_\beta$, $v_{\gamma\delta}b_\gamma^+b_\delta$,
$h_{\alpha\beta\gamma\delta}a_\alpha^+a_\beta b_\gamma^+b_\delta$
and their hermitian conjugate. The operation
$a_\kappa^+a_\lambda b_\mu^+b_\nu \mid\{n_\alpha\}\{n_\beta\}\rangle$
is defined by eq. (\ref{prod}).

\subsubsection{Telegraph signals ("quantum jumps")}\label{telegraph}
Whereas the theory is relativistic, i.e., Lorentz invariant, the interesting
solutions emerge in the non-relativistic limit.
If the time dependent wave functional is calculated along the lines just
outlined,
one finds that configurations $\mid\{n_\alpha =1\}\{n_\beta =0\}\rangle$
restricted to four dimensional spacetime ("three dimensional
configurations")
become heavily entangled with configurations living essentially in eleven dimensional
spacetime.
The reason is that due to the high density of "gravonon configurations"
(i.e. configurations where gravonons in the hidden dimensions have been excited, cf. section \ref{gravononsect})
a particular four dimensional configuration will always be practically degenerate with
gravonon configurations and, as it is well known, the \SE will mix
degenerate configurations
even for vanishingly small interaction strength.\\

\noindent
In the problem of diffusion of an adsorbed particle on a solid surface the Hamiltonian has been chosen as
\cite{doyenDICE2012,drakovaDICE2012}:
\begin{eqnarray} \label{hamiltonian}
{\rm H}&=&{\rm H}_{gas\;atom}+{\rm H}_{graviton}+{\rm H}_{gas\;atom-graviton}\nonumber \\
&=&E_{g_1}n_{g_1} +E_{g_2}n_{g_2}  + E_{w_1}n_{w_1}+E_{w_2}n_{w_2} +
\sum_{i=1}^2V_{loc}^{g_i}(a^+_{g_i}a_{w_i} + a^+_{w_i}a_{g_i})  \nonumber \\
&+& \sum_{i=1}^2\{\varepsilon_{grav_i}b^+_{grav_i}b_{grav_i} +\sum_{k}
\varepsilon_{k_i} b^+_{k_i}b_{k_i} \nonumber \\                                                     
&+& \sum_{k} [V_{grav,w_i}n_{w_i}b^+_{grav_i}b_{k_i}+
V_{w_i,grav_i}n_{w_i}b^+_{k_i}b_{grav_i}]\} .
\end{eqnarray}
The meaning of the symbols is:
$n_{g_i}, n_{w_i}$: gas particle field strength in the
vibrationally ground and excited (parallel to the surface)
core movement states and in the "warp resonance" where interaction with
the gravonons occurs;
$a^+_i, a_j$: creation and annihilation operators for the gas particle field
in the respective core movement states;
$V_{loc}^{g_i}$: interaction strength between a gas particle
state and the warp resonance;
$\varepsilon_{grav_i},\varepsilon_{k_i}$: energy quanta of local and continuum
gravonon fields;
$b^+_{grav_i}, b_{grav_i}$: creation and annihilation operators for
the local gravonon field;
$b^+_{k_i}, b_{k_i}$: creation and annihilation operators for
the continuum gravonon field;
$V_{grav,w_i}$: interaction between the gas particle and the gravonons
within the warp resonance.\\

\noindent
The field configurations taken into account are the following:
\begin{equation}
\{n_\alpha\}= n_{g1}n_{g2}n_{w1}n_{w2}
\end{equation}
with $n_{gi}=0,1$, $n_{wi}=0,1$ for $i=1,2$.
\begin{equation}
\{n_\beta\}= n_{k1}n_{k2}
\end{equation}
with $n_{ki}=0,1,...,\infty$ for $i=1,2$.\\

\begin{figure}{} \begin{center} \begin{minipage}{11cm}
\scalebox{0.5}{\includegraphics*{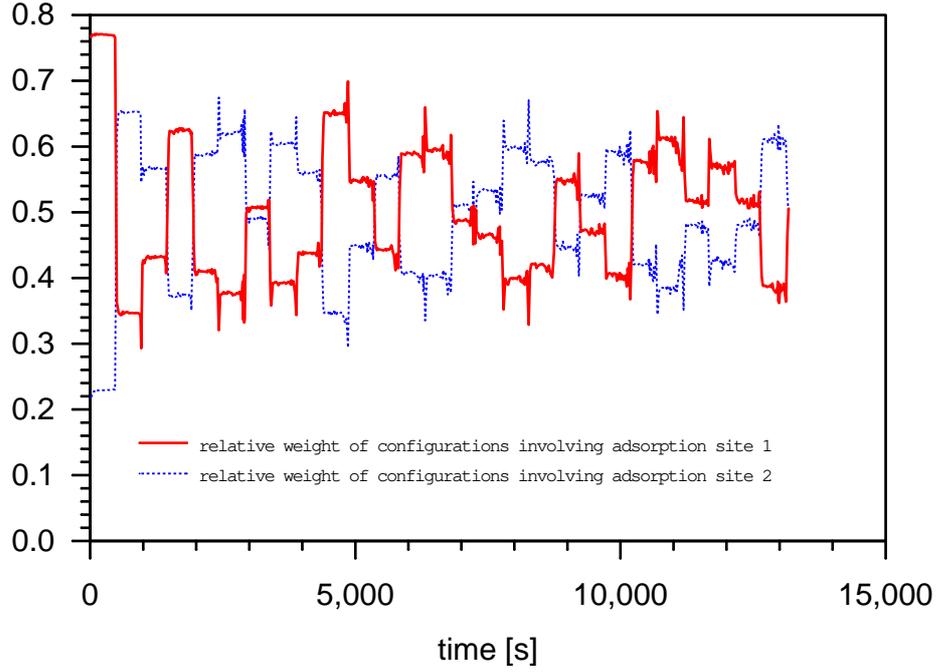}}
\end{minipage}
\end{center}
\caption{Adsorbate telegraph-signal-like quantum jumps 
between two adsorption sites as a function of time
obtained as a solution of the \SE .
The entanglement with the gravonon field is included.
The quantity plotted is the squared coefficient of the field configurations containing gravonons
summed over site 1 and site 2, respectively, with a full and a dashed curve.
\label{telegraph-jumps}} \end{figure}

\noindent
The solution of the \SE for this Hamiltonian
results in
the adparticle being fixed near a particular substrate atom as long as
the excited gravonons are moving around in the large hidden dimensions. If
after the recurrence time the gravonons return to the four dimensional brane where they
have been excited, the adparticle is suddenly free to leave this site.
When arriving
at a neighbouring site the adparticle will excite gravonons here and will be
captured to this new site as long as the excited gravonons are moving
around
in the large hidden dimensions. This leads to the telegraph signal like
adsorbate movement, which is observed in experiment. A calculation of
this movement
within the present theory is depicted in fig. \ref{telegraph-jumps}.
The analytical derivation of why and how telegraph signal like  
changes in the state of a quantum system, entangled with 
gravonons in the environment, occur, is presented in ref. \cite{arxive}.

\subsubsection{The chooser mechanism ("collapse")}\label{choosermod}
The scattering of a matter field from a double slit and the
"wave - particle duality" observed in these experiments is in our theory interpreted
as follows.
If there is a site on the screen which is energetically degenerate with
the initial wave functional $\mid Q0\rangle =\mid\Psi (t=0)\rangle$ 
the configuration $\mid K\kappa_{proj}\rangle$ with the
scalar matter field $\phi$ having significant strength on this site will get strongly entangled
with the degenerate gravonon configurations. As the excitations of the gravonons
in the hidden dimensions depend on the matter field being extremely
localized near the chosen site the matter field strength cannot drift away as long as
the excited gravonons are moving around in the large hidden dimensions.
If this takes time long enough for other physical-chemical processes to be
initiated, it will result in an experimentally detectable event.
Such an event would be called "collapse" in the framework of Copenhagen quantum mechanics. \\

\begin{figure}{} \begin{center} \begin{minipage}{11cm}
\scalebox{0.5}{\includegraphics*{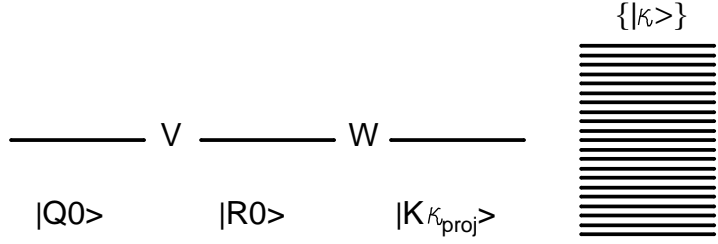}}
\end{minipage}
\end{center}
\caption{Field configurations and their interactions: 
$\mid Q0 \rangle=\mid Q \rangle \otimes\mid 0_{grav} \rangle$ initial field configuration with the 
gravitational field in the ground state and the matter field in the source; 
$\mid R0 \rangle=\mid R \rangle \otimes\mid 0_{grav} \rangle$ 
with the matter field localized on the screen but
not yet entangled with the gravonons, which are in the ground state;    
$\mid K\kappa_{proj} \rangle$ with the matter field localized on the screen and
a gravonon state projected out of the gravonon band $\{ \mid \kappa \rangle \}$  
according to the criterion of strongest interaction.
\label{QRK-states}} \end{figure}

\noindent
To treat this situation theoretically in a simplified manner
we set up the Hamiltonian ${\rm H}$ in matrix form
\begin{center}
\begin{tabular}{c|ccc}
  & $\mid Q0\rangle$     & $\mid R0\rangle$ & $\mid K\kappa_{proj}\rangle$ \\
\hline  
$\langle Q0\mid$ & $0$ & $V$ & 0  \\
$\langle R0\mid$ & $V$ & 0 & W  \\
$\langle K\kappa_{proj}\mid$ & 0 & W & 0
\end{tabular}
\end{center}
Figure \ref{QRK-states} illustrates the many-particle states involved and their 
interactions.
$\mid R0 \rangle=\mid R \rangle \otimes\mid 0_{grav} \rangle$ 
is a state where the matter field is localized on the screen but
not yet entangled with the gravonons. In the state $\mid K\kappa_{proj}\rangle$
the matter field experiences
the interaction with the gravitational field. 
All three states are degenerate at energy
$E=0$ which defines the energy zero. The interaction $V$ allows the matter field to penetrate
into the localized state $\mid R0 \rangle$ on the screen and $W$ is the potential
connecting $\mid R0 \rangle$  to the resonance $\mid K\kappa_{proj} \rangle$
where gravitational interaction becomes important.\\

\noindent
Diagonalizing the Hamiltonian leads to the characteristic equation
\begin{equation}
{\rm det}(E-{\rm H}) = E^3-(W^2+V^2)E =0
\end{equation}
which has the solutions $E_o=0$, $E_\pm =\pm \sqrt{W^2+V^2}$. Only the eigenstate $\mid 00\rangle$
of energy zero is of interest in the following. It has the coefficients
\begin{eqnarray}
C_{Q0}&=&\frac{W}{\sqrt{V^2+W^2}}\\
C_{R0}&=&0\\
C_{K\kappa_{proj}}&=&\frac{V}{\sqrt{V^2+W^2}}.
\end{eqnarray}
In the limit where the coupling to the gravonons becomes very weak ($\frac{W}{V}\rightarrow 0$)
$C_{Q0}=0$ and $C_{K\kappa_{proj}}=1$. The coefficients of the "zero-state" $\mid 00\rangle$ in the eigenstates
of the total system become then approximately
\begin{eqnarray}
\langle 00\mid K\kappa+\rangle &=&\langle 00\mid {\rm G}^+(\epsilon_{\kappa})(W_{grav}+V+W)\mid K\kappa\rangle\\
&\approx & \langle K\kappa_{proj}\mid {\rm G}^+(\epsilon_{\kappa})(W_{grav}+V+W)\mid K\kappa \rangle\\
&=&\langle K\kappa_{proj}\mid {\rm G}^+(\epsilon_{\kappa})\mid K\kappa_{proj}\rangle\langle K\kappa_{proj}\mid W_{grav}\mid K\kappa\rangle\\
&=&G_{K\kappa_{proj}}^+(\epsilon_{\kappa})W_{\kappa}
\end{eqnarray}
where ${\rm G}^+$ is the Green operator, $W_{grav}$ is the gravitational potential,
$G_{K\kappa_{proj}}^+=\langle K\kappa_{proj}\mid {\rm G}^+\mid K\kappa_{proj}\rangle$ and 
$W_{\kappa}=\langle K\kappa_{proj}\mid W_{grav}\mid K\kappa\rangle$.
$\mid \kappa \rangle$ is one of the states in the gravonon band, generating 
$\mid \kappa+ \rangle$ after the interaction with the matter field is included. 
$\epsilon_{\kappa}$ is the eigenenergy of the eigenstate $\mid Kk+\rangle$.
We have here applied the Lippmann-Schwinger equation \cite{lippschwi} to construct
the eigenstates $\mid K\kappa+\rangle$. The Green function $G_{K\kappa_{proj}}^+$ has the form
\begin{equation}
G_{K\kappa_{proj}}^+(\epsilon_{\kappa})=\frac{1}{\epsilon_{\kappa}-\alpha(\epsilon_{\kappa})+i\Gamma(\epsilon_{\kappa})}
\end{equation}
where $\alpha$ and $\Gamma$ are the real and imaginary parts of the self-energy,
respectively, with 
\begin{eqnarray} \label{kappaproj}
\Gamma (\epsilon_k )&=&\pi \sum_{\kappa}\mid W_{\kappa}\mid^2\delta (E-\epsilon_{\kappa})\\
&=&\pi\frac{U^2}{\Delta }\;\; {\rm i}{\rm f} \mid\epsilon_k\mid < \Delta
\end{eqnarray}
The last line indicates an approximation where $W_{\kappa}=\frac{U}{\sqrt{N}}$ is non-zero and energy independent
over an energy range of length $\Delta $ and zero otherwise. $N$ is the number of gravonons in the
energy interval $\Delta$.
Defining the density $\rho_{\epsilon}$ of gravonon levels at $\epsilon_{\kappa}=0$ 
\begin{eqnarray}\label{rhoeq40}
\frac{\rho_\epsilon}{N}&=&\frac{1}{\Delta}\\
&=&-\frac{1}{\pi}{\rm I}{\rm m}{\rm G}^+_{K\kappa_{proj}}\\
&=&\frac{1}{\pi\Gamma} 
\end{eqnarray}
yields $\Delta = \pi\Gamma$. 
In order to study the time dependence due to the coupling to the gravonons
we start from the formal solution of the time dependent \SE
\begin{eqnarray}   \label{psioft}
\mid\Psi (t)\rangle &=& e^{-i{\rm H}t}\mid 00\rangle \nonumber \\
&=&\sum_{\kappa}{\rm e}^{-i\epsilon_{\kappa}t}\mid K\kappa+\rangle\langle K\kappa+\mid 00\rangle  \nonumber \\
&=&\sum_{\kappa}{\rm e}^{-i\epsilon_{\kappa}t}\mid K{\kappa}+\rangle G_{K\kappa_{proj}}^-(\epsilon_{\kappa})W_{\kappa}  \nonumber \\
&=&\sum_{\kappa}{\rm e}^{-i\epsilon_{\kappa}t}\mid K{\kappa}+\rangle \frac{W_{\kappa}}{\epsilon_{\kappa}-\alpha(\epsilon_{\kappa})-i\Gamma(\epsilon_{\kappa})}\label{time00}
\end{eqnarray}
where we assumed $W_{\kappa}$ to be real. 
For the projection $\langle K\kappa_{proj} \mid \Psi (t)\rangle$ we obtain: 
\begin{equation} \label{psiprojcted}
\langle K\kappa_{proj}\mid\Psi (t)\rangle =
\sum_{\kappa}{\rm e}^{-i\epsilon_{\kappa}t} \langle K\kappa_{proj} \mid K\kappa+\rangle \frac{W_{\kappa}}{\epsilon_{\kappa}-\alpha(\epsilon_{\kappa})-i\Gamma(\epsilon_{\kappa})}.
\end{equation}
From eq. (\ref{kappaproj}), with $W_{\kappa}$ assumed to be independent of $\epsilon_\kappa$,
we have:
\begin{equation}
\langle K\kappa_{proj}\mid  K{\kappa}+\rangle W_{\kappa}=\frac{U}{N}.
\end{equation}
Inserting in eq. (\ref{psiprojcted}) we obtain
\begin{equation}\label{time01}
\langle K\kappa_{proj}\mid\Psi (t)\rangle 
=\frac{U}{N} \sum_{\kappa}\frac{{\rm e}^{-i\epsilon_{\kappa}t}}{\epsilon_{\kappa}-\alpha(\epsilon_{\kappa})-i\Gamma(\epsilon_{\kappa})}
\end{equation}
where $\langle K\kappa_{proj}\mid K{\kappa}+\rangle=\frac{1}{\sqrt{N}}$ is 
assumed to be independent of $\epsilon_{\kappa}$.
\noindent
We have to emphasize at this point that the summation index $\kappa$
is {\em not} the graviton wave vector. The sum runs over the energy levels of the gravonon states
(cf. section \ref{gravononsect}).
The sum in eq. (\ref{time01}) can therefore be transformed into an integral by writing
\begin{equation}
\langle K\kappa_{proj}\mid\Psi (t)\rangle 
=\frac{U}{N} \int_{-\infty}^\infty {\rm d}\epsilon_{\kappa}\rho_\epsilon\frac{{\rm e}^{-i\epsilon_{\kappa}t}}{\epsilon_{\kappa}-\alpha(\epsilon_{\kappa})-i\Gamma(\epsilon_{\kappa})}.
\end{equation}
($\rho_\epsilon$, the level density of gravonon states, is constant according to eq. (\ref{rhoeq40}).) We assume
$\alpha$ and $\Gamma$ to be independent of energy.
The time dependent wave functional $\mid\Psi (t)\rangle$ and its component 
involving the localized matter state $\mid K \rangle$ is then obtained by
integration yielding \cite{gradshteyn}
\begin{equation}\label{psiproj}
\langle K\kappa_{proj}\mid\Psi (t)\rangle =
i \pi \frac{U}{N} \rho_\epsilon {\rm e}^{-\Gamma t} =
i \pi \frac{U}{\Delta}{\rm e}^{-\Gamma t},
\end{equation}
where in the weak interaction limit we assumed $\alpha \rightarrow 0$ and 
$\rho_\epsilon=\frac{N}{\Delta}$ (eq. \ref{rhoeq40}).\\

\noindent
In order to clarify the complete character of the state $\mid\Psi (t)\rangle$ we have also to calculate
its projections on $\mid R0\rangle$, $\mid Q0\rangle$ and $\mid K\lambda +\rangle$.
\begin{eqnarray}
\langle R0\mid K\kappa +\rangle &=& \langle R0\mid G^o(W_{grav}+V+W)\mid K\kappa +\rangle\\
&=&G^o_{R0}\langle R0\mid (W_{grav}+V+W)\mid K\kappa +\rangle\\
&=&G^o_{R0}\langle R0\mid (W_{grav}+V+W)\mid K\kappa_{proj}\rangle\langle K\kappa_{proj}\mid K\kappa +\rangle\\
&=& \frac{W}{\sqrt{N}}\lim_{\epsilon\rightarrow 0}\frac{1}{\epsilon_\kappa +i\epsilon}\\
&=& \frac{W}{\sqrt{N}}\left [{\cal P}(\frac{1}{\epsilon_\kappa})-i\pi\delta(\epsilon_\kappa ) \right ]
\end{eqnarray}
Using this in eq. (\ref{psioft}) yields then with $W_k=U/\sqrt{N}$ and $\rho_\epsilon =N/\Delta$
\begin{equation}
\langle R0\mid \Psi (t)\rangle = \pi\frac{UW}{\Delta(\Gamma -i\alpha ) }
\end{equation}
\begin{eqnarray}
\langle Q0\mid K\kappa +\rangle &=& \langle Q0\mid G\mid K\kappa_{proj}\rangle
\langle K\kappa_{proj}\mid W_{grav}\mid K\kappa \rangle\\
&=&\frac{U}{\sqrt{N}}\langle Q0\mid G\mid K\kappa_{proj}\rangle
\end{eqnarray}
Inserting the Born series
\begin{equation}
G=G^o+\sum_{n=0}^\infty (G^o(V+W)G^o)^n
\end{equation} 
one has
\begin{eqnarray}
\langle Q0\mid K\kappa +\rangle &=& 
\frac{W}{\sqrt{N}}\sum_{n=0}^\infty
(\langle Q0\mid G^o\mid Q0\rangle\langle Q0\mid V\mid R0\rangle\langle R0\mid G^0\mid K\kappa_{proj}\rangle )^n
\end{eqnarray}
which is zero at every order and hence $\langle Q0\mid K\kappa +\rangle =0$.
Requiring unitarity and setting $\alpha =0$ one gets
\begin{eqnarray}
\sum_\lambda\mid\langle K\lambda +\mid \Psi (t)\rangle\mid^2 &=& 
1-\mid\langle K\kappa_{proj}\mid\Psi(t)\rangle\mid^2-\mid\langle R0\mid\Psi(t)\rangle\mid^2
-\mid\langle Q0\mid\Psi(t)\rangle\mid^2 \nonumber \\
&& \\
&=& 1-\pi^2 \left [\frac{U^2}{\Delta^2}e^{-2\Gamma t}+(\frac{UW}{\Delta\Gamma})^2+0 \right ] \\ \label{chooserana}
&\approx &1-e^{-2\Gamma t}-\frac{W^2}{U^2}
\end{eqnarray}
The last line is valid, because from eq. (\ref{kappaproj}) $\frac{U^2}{\Delta}=\Gamma /\pi$ and $\Delta = \pi\Gamma$
(eq. \ref{rhoeq40}).\\
\begin{figure}{} \begin{center} \begin{minipage}{11cm}
\scalebox{0.5}{\includegraphics*{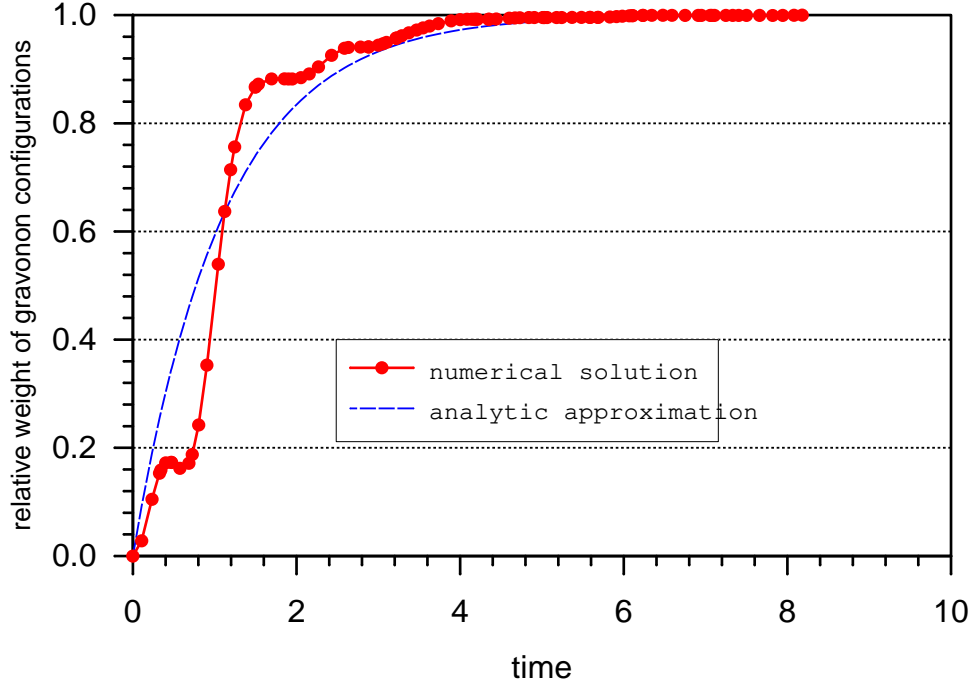}}
\end{minipage}
\end{center}
\caption{The chooser at work: the numerical solution of the model of section \ref{choosermod}
in the basis displayed in fig. \ref{QRK-states} (full curve)
in comparison to the analytical approximation eq. (\ref{chooserana}) (dashed curve).
\label{chooserfig}} \end{figure}

\noindent
Neglecting off-shell components, the initial state $\mid Q0\rangle$, by propagating through the nano lattice (or double slit),
becomes the state $\mid 00\rangle$, which for weak gravitational interaction strength merges into the
state $\mid K\kappa_{proj}\rangle$ and then
is completely extinguished with time according to eq. (\ref{psiproj}). 
The whole field strength is captured in the states $\{\mid K{\kappa}+\rangle\}$,
representing a localized matter field, strongly entangled with gravonon states, which live mainly in the hidden
dimensions.\\

\noindent In fig. \ref{chooserfig} the sum of the weights of the gravonon configurations
(squared coefficients) in the wave functional shows that as the system, displayed 
in fig. \ref{QRK-states}, develops with time the weight of the field configurations 
with gravonon components increases, tending to 1. This means that the matter field  
is localized and a ''collapse'' occurs on a single site $\mid K \rangle$ where the  
entanglement with the gravonons is effective. The oscillations in the exact 
numerical calculation arise, as the off-shell states with $E_{\pm}$ are also mixed in 
because of the finite energy spread of the gravonons. \\

\noindent
In the present theory the experimentally detectable event is uniquely
determined already at the moment of the emission of the matter field quantum from
the source, because exact energetical degeneracy is needed and, as outlined in section
\ref{hidden}, due to the weakness of the interaction there is only one site
of suitable energy available on the screen.

\subsubsection{Beables and non-locality}
Following Bell \cite{bell,Bell2} we exclude the notion of observables
in favour of beables.
The beables \cite{bell,Bell2} of our theory are the field configurations where
matter
fields extremely localized in three dimensional space are entangled to gravonons
moving in
11 dimensional spacetime. These beables emerge in the non-relativistic limit
within a preferred Lorentz frame, corresponding to the rest frame of the
investigated system, the laboratory and the experimenter.\\

\noindent
Unlike \CQM where few "observables" have to be chosen from an
infinite set of hermitian operators by using criteria which lie outside
the realm of Copenhagen quantum mechanics,
namely by being found suitable  for the interpretation of experiments,
the beables
which define measurement are uniquely defined within our theory, based on the Schr\"odinger
equation.\\

\noindent
It is interesting that our definition of a beable is compatible with
Kant's ideas
about observations \cite{kant}. According to Kant, knowledge rests on
our sensual
contact with the world, and our sensibility has its limits. The sensual
limits
Kant called the {\em forms of intuition}. He argued that the forms are
space and time \cite{auyang}.\\

\noindent
The intuitive nature of our beables is in marked contrast to the
artificial constructs
which are sometimes necessary in \CQM to define e.g. the measured
expectation values of the electromagnetic field (see \cite{auyang}, note
43). \\

\noindent
The transition from configuration
space to the three dimensional space experienced by the experimenter is unambiguously
provided by the localization in \trdim space of the matter field.
As the beables emerge (in the non-relativistic limit) from a solution of
the \SE in the (high-dimensional) configuration space, the beables, as
observed
in three dimensional space,
clearly exhibit non-locality, i.e., beables can appear simultaneously at
different points
in \trdim space, if we describe the \trdim space  in the rest frame.\\

\noindent
The fact that the beables only emerge in the non-relativistic limit and
are neither
defined nor empirically accessible for relative velocities (between
experimenter
and physical event), approaching the velocity of light, resolves the apparent paradox of
"real collapses in Minkowski spacetime" \cite{callender}.
According to this paradox beables which appear simultaneously in one
Lorentz frame appear at different times in another Lorentz frame.
This difference in time is noticeable only, if the relative velocity
of the two Lorentz frames approaches the velocity of light.\\

\noindent
As both the exerimenter and the physical event which she is to observe
consist of beables, according to the present theory the beables
representing "collapses" and/or "quantum jumps" are not empirically
accessible as such by the experimenter, if the physical event moves with
a velocity
near to the \vl relative to the experimenter. This is associated with
the fact that
our theory does not contain "collapses" or "quantum jumps" as additional
processes compared to Copenhagen quantum mechanics, where these processes occur in all Lorentz
frames immediately. In our theory collapses are replaced by deterministic
localization processes which can occur and switch in a \tsl way.
These \tsl processes occur in the rest frame of the experiment within
a very short time, but they are not immediate. If the observer moves
with a high velocity relative to the experimental setup, the \tsl processes
slow down and can no longer represent "collapses" or "quantum jumps"
of the \CQM type.\\

\noindent
This is the well established time dilatation which is present in our
Lorentz
invariant theory. As is common in popular presentations of special
relativity this can be described in the way that for high relative velocity
the observer experiences an increased mass of the matter and
gravonon fields which leads to a slower movement. Processes,
which in the rest frame of the experimental setup appear immediate
and simultaneous, are realized in a boosted Lorentz frame as blurred
in time and simultaneity of the experiments is no longer defined.\\

\noindent
We can elaborate on this aspect in the following way. Assume that Alice
and Bob perform an EPR-experiment \cite{EPR} in earth bound laboratories and
measure at distant locations but simultaneously the spins of two particles
bound in a singlet state, thus verifying the established non-locality
emanating from the Schr\"odinger equation. At the moment when they measure the
spins (i.e. simultaneously) each one sends a light signal to a rocket,
which moves with a velocity near to the \vl above them in the direction
from Bob to Alice. The observer in the rocket registers the light signal
from Alice earlier than that from Bob and therefore concludes in the
rocket bound
rest frame that the two measurements have not been performed simultaneously.\\

\noindent
Why does the apparent paradox in Minkowski space exist?
This is the kind of reasoning that leads to it. The observer in the
rocket would say that in the time
interval, which starts, when she "sees" that Alice measures the spin of
particle 1,
and ends, when she "sees" that Bob measures the spin of particle 2,
the spin of particle 2 is not determined. In this time interval it is
therefore not
guaranteed that the system, consisting of the two particles, is in a
singlet state, i.e.,
that the total spin is conserved. This means, the observer in the rocket
finds
that the rules of \CQM are violated (cf. e.g. \cite{tooley}).
The Schr\"odinger equation, however, requires that there can be no temporal gap between
the measurement of the spin of particle 1 and the time, when the spin of
particle 2
acquires a definite value. One further reasons that the rest frame of
the earth
and of the rocket are connected by a Lorentz transformation and, hence, are
fully equivalent. It is deduced then that the two events (measuring the
spin of particle 1 and measuring
the spin of particle 2) are real and immediate and their existence is
independent of the chosen
Lorantz frame. However, the existence of an event in the rocket, (i.e.
the recording of the photons in the
rocket bound reference frame), and events in the earth bound reference
frame,
(i.e. the measurement of the spin of particle 1 by Alice simultaneously
with the measurement of
the spin of particle 2 by Bob), are associated with different beables,
hence with
further events. The fact that events additional and different from
Alice's and Bob's
measurements have to occur (the light signals hitting the retina or some
registration device and initiating a physical-chemical process) is not
considered
to be of importance and is not discussed.\\

\noindent
In our theory, however, every measurement is a beable of the kind defined
with the first sentence in the present chapter.
If the rocket stays far off the
laboratories on earth, the judgement of the observer in the rocket
has to rely on the information contained in the
light signal as born out by the beables it generates in the rocket.
The light emission process which is immediate in the rest frame
of the experimental setup appears extended in time in the rocket.\\

\noindent
The reasoning presented here is only valid, because in the first place
the beables do
not emerge independent of the reference frame and in the second place
measurement is a consequence of the Schr\"odinger equation. In our theory all events, i.e.
beables,
in the rest frame of the rocket (including those constituting the
observer) emerge from the \SE in
the rest frame of the observer. Therefore, there is no paradox, i.e., no
contradiction
either to the observed facts, or to the symmetry of Lorentz invariance,
or to the predictions of the Schr\"odinger equation.\\

\noindent
In our theory the wave functional $\mid \Psi\rangle$ yields a complete
description of the world. The beables, which constitute both observables
and measurements, are generated by solving the \SE for the wave
functional in the
non-relativistic limit. All measurements
of various physical quantities can be traced back to determining the
location of the particles \cite{albert}.

\subsubsection{A remark on entropy}
Our world wave functional is unique and a solution of the Schr\"odinger equation.
Therefore the
von Neumann entropy associated with this wave functional is zero. Entropy,
however, in its statistical and /or thermodynamic applications is clearly
a quantity which should be defined in four dimensional spacetime. As the events in
four dimensional
spacetime are entangled to events in the seven hidden dimensions, an
evaluation of entropy in four dimensional spacetime would be conventionally tried
by using the entanglement
entropy.\\

\noindent
To define entropy in our theory by means of the density matrix is, however,
not possible, because $\mid \Psi \rangle$ does not have the meaning of a probability
amplitude. It is rather a description of the various fields (matter,
electromagnetic,
gravitational) in configuration space. A configuration
$\mid \{ n_\alpha\}\{ n_\beta \}\rangle $
is a distribution of the fields in 10 dimensional space.\\

\noindent
Entropy has then to be defined as being proportional to the
logarithm of the number of microscopic field configurations, representing
a macroscopic field configuration. In our case this is then the logarithm
of the number of initial conditions (source of matter field,
number of gravonon structures $\{ n_{\beta} \}$) which lead to the state
$\mid n_{\alpha} \rangle$ for the matter field in four dimensional spacetime.\\

\noindent
Compared to \CQM we switch from a probabilistic interpretation of
$\mid\{n_\alpha\}\rangle$ to a statistical interpretation,
i.e., the number of ways a "three dimensional beable" can be realized by
"10 dimensional beables". Probability implies that we have a principal
knowledge gap, i.e., it is principally impossible for us to know,
if the "three dimensional beable" will be realized. Statistics means that we can
predict the occurrence of the "three dimensional beable", if we start from
a precisely defined initial condition. Due to limited experimental
resolution we have a sample of many initial conditions which
permits only statistical statements.\\

\noindent
Clearly the entropy increases steadily as further beables are
created.
The entanglement between the matter fields and the gravonons persists forever,
because there is nothing like a "collapse" of the total wave
functional
or any other external influence. The total (world) wave functional develops
continuously and deterministically according to the dynamics dictated by
the Schr\"odinger equation. The second law is thus a consequence of our theory and need not
be introduced as an additional postulate.

\section{Effective Hamiltonian for matter-graviton interaction}
\noindent In this section the effective Hamiltonian of our Emerging quantum mechanics 
is derived from first principles to describe the matter-graviton interaction. 
The Hamiltonian describes the matter field as a scalar
boson field (denoted $\phi$ later on) in interaction with a gravonon
field (denoted $\zeta_{00}$ later on) in the presence of 
gravitational interaction. This derivation
shows how the gravonon field emerges from linearized general relativity.

\subsection{Weak field gravity}
\noindent
The following introduction to  gravity as a field theory is in the spirit
of Zee \cite{zee}. Units are $\hbar = {\rm c} =1$.\\

\noindent
The Einstein-Hilbert action for gravity in $N$ spacetime dimensions is
\begin{equation} \label{einsteinhilbert}
S_{grav}=\frac{1}{16\pi G^{(N)}}\int {\rm d}^Nx\sqrt{-g}R,
\end{equation}
where $g={\rm det}g_{\mu\nu}$ denotes the determinant of the curved metric $g_{\mu\nu}$
of spacetime, $R$ is the scalar curvature, and $G^{(N)}$ is the gravitational constant
in $N$ dimensional spacetime. The sign convention varies widely in the literature \cite{misner},
we use the sign convention of \cite{zee}.\\

\noindent
In addition to gravity there are matter fields 
embedded in curved spacetime, for instance the electromagnetic field
which lives only in four dimensional spacetime,
as contributions $S_{matter}$ to the total action:
\begin{equation}
S_{tot}=S_{grav}+S_{matter}
\end{equation}
$S_{tot}$ is assumed to describe the real world, if the theory is properly quantized.
Quantization is only possible for weak gravitation. This is the procedure adopted
here.\\

\noindent
Alternatively one could treat quantum fields in a classical gravitational background.
This is, however, a semi-classical approximation, which does not define a unique
vacuum state and/or a unique world wave function. It would not allow to describe
the entanglement of matter fields with gravitons which is proposed to contribute
in a significant way to effects observed in experiment.
Therefore the semi-classical theory is not followed here.\\

\noindent
The stress-energy tensor of the matter fields is derived from the action $S_{matter}$:
\begin{eqnarray} \label{stressenergy}
T^{\mu\nu}(x)&=&-\frac{2}{\sqrt{-g}}\frac{\delta S_{matter}}{\delta g_{\mu\nu}(x)} \nonumber\\
T_{\mu\nu}(x)&=&-\frac{2}{\sqrt{-g}}\frac{\delta S_{matter}}{\delta g^{\mu\nu}(x)}.
\end{eqnarray}
$x$ is a $N$-dimensional vector denoting a point in $N$ dimensional spacetime.
The signature is $(+,-,-,-,-,-,-,-, ...,-)$. These equations are valid in general.
The weak field limit is defined by a linear deviation from 
the Minkowski metric $\eta_{\mu\nu}$:
\begin{equation} \label{lineargrav}
g_{\mu\nu}=\eta_{\mu\nu}+h_{\mu\nu}.
\end{equation}
Our philosophy is that this form of the metric is the correct one for
the physics we want to investigate.
Expanding $S_{matter}$ to first order in $\delta g_{\mu\nu} = h_{\mu\nu}$
\begin{equation}
S_{matter}(h_{\mu\nu})=S_{matter}(h_{\mu\nu}=0)+\delta S_{matter}
\end{equation}
and plugging in $\delta S_{matter}$ from eq. (\ref{stressenergy})
one obtains:
\begin{eqnarray} \label{interactiongeneral}
\delta S_{matter}&=&-\int {\rm d}^Nx\sqrt{-g} \frac{1}{2}h_{\mu\nu}T^{\mu\nu}\nonumber\\
&=&-\int {\rm d}^Nx\sqrt{-g} \frac{1}{2}h^{\mu\nu}T_{\mu\nu}.
\end{eqnarray}
$\sqrt{-g}$ is expanded as
\begin{equation} \label{sqrt-g}
\sqrt{-g} \approx 1+\frac{\partial\sqrt{-g}}{\partial h^{\mu\nu}}h^{\mu\nu}.
\end{equation}
Using the following general relationship (Palatini method) 
\begin{equation}
\frac{\partial\sqrt{-g}}{\partial g^{\alpha\beta}}=-\frac{1}{2}\sqrt{-g}g_{\alpha\beta}
\end{equation}
together with
\begin{equation}
\frac{\partial\sqrt{-g}}{\partial g^{\alpha\beta}}=\frac{\partial\sqrt{-g}}{\partial h^{\alpha\beta}}
\end{equation}
and $\sqrt{-g}=1$, we obtain:
\begin{eqnarray}
\frac{\partial\sqrt{-g}}{\partial h^{\alpha\beta}}&=&-\frac{1}{2}\left( 
1+\frac{\partial\sqrt{-g}}{\partial h^{\mu\nu}}h^{\mu\nu}\right)(\eta_{\alpha\beta}+h_{\alpha\beta})\\
&=& -\frac{1}{2}\eta_{\alpha\beta} -\frac{1}{2}h_{\alpha\beta}-\frac{1}{2}\frac{\partial\sqrt{-g}}{\partial h^{\mu\nu}}h^{\mu\nu}\eta_{\alpha\beta}
-\frac{1}{2}\frac{\partial\sqrt{-g}}{\partial h^{\mu\nu}}h^{\mu\nu}h_{\alpha\beta}\\
&\approx& -\frac{1}{2}\eta_{\alpha\beta} -\frac{1}{2}h_{\alpha\beta}
+\frac{1}{4}(\eta_{\mu\nu}+h_{\mu\nu})h^{\mu\nu}\eta_{\alpha\beta}
+\frac{1}{4}(\eta_{\mu\nu}+h_{\mu\nu})h^{\mu\nu}h_{\alpha\beta}.
\end{eqnarray}

\noindent Omitting terms of higher order than linear in $h_{\alpha\beta}$ and plugging this into
eq. (\ref{sqrt-g}) yields
\begin{equation}
\sqrt{-g}=1-\frac{1}{2}\eta_{\mu\nu}h^{\mu\nu}-\frac{1}{2}h_{\mu\nu}h^{\mu\nu}
+\frac{1}{4}\eta_{\alpha\beta}h^{\alpha\beta}\eta_{\mu\nu}h^{\mu\nu}.
\end{equation}
Inserting this in eq. (\ref{interactiongeneral}) we obtain, neglecting the last two terms
in the previous equation:
\begin{eqnarray} \label{interaction}
\delta S_{matter}&=&-\int {\rm d}^Nx\frac{1}{2}h_{\mu\nu}T^{\mu\nu}(1-\frac{1}{2}\eta_{\alpha\beta}h^{\alpha\beta}).
\end{eqnarray}
A matter field is any field that is not the graviton field.
Imposing the harmonic gauge condition (cf. refs. \cite{zee} and \cite{ryder})
\begin{equation} \label{harmonicgauge}
\partial_\mu h^\mu_\nu = \frac{1}{2}\partial_\nu h^\lambda_\lambda
\end{equation}
($h_{\nu}^{\mu}=h_{\nu\alpha}g^{\alpha\mu}$)
the action for the weak field gravity becomes (ref. \cite{zee}):
\begin{eqnarray} \label{weakfieldgravity}
S_{wfg}&=&S_{grav}(g_{\mu\nu}=\eta_{\mu\nu}+h_{\mu\nu})+\delta S_{matter} \nonumber \\
&=&\int {\rm d}^Nx\frac{1}{2}\left[ \frac{1}{32\pi G^{(N)}}(\partial_\lambda h^{\mu\nu}\partial^\lambda h_{\mu\nu}
-\frac{1}{2}\partial_\lambda h^\mu_\mu\partial^\lambda h^\mu_\mu )-h_{\mu\nu}T^{\mu\nu}(1-\frac{1}{2}
\eta_{\alpha\beta}h^{\alpha\beta})\right]. \nonumber \\
\end{eqnarray}
The total action is then:
\begin{equation}
S_{tot}=S_{wfg}+S_{matter}(h_{\mu\nu}=0).
\end{equation}

\noindent Varying eq. (\ref{weakfieldgravity}) with respect to $h_{\mu\nu}$ we obtain
the textbook result for the
Euler-Lagrange equation of motion (ref. \cite{zee}):
\begin{equation} \label{eulerlagrange}
\partial^2 h_{\mu\nu} = -16\pi G^{(N)}(\tilde{T}_{\mu\nu}
- \frac{1}{2}\eta_{\mu\nu}\tilde{T})
\end{equation}
where we define:
\begin{eqnarray}
\tilde{T}_{\mu\nu}&=&T_{\mu\nu}(1-\frac{1}{2}\eta_{\alpha\beta}h^{\alpha\beta})\\
\tilde{T}&=&T(1-\frac{1}{2}\eta_{\alpha\beta}h^{\alpha\beta})
\end{eqnarray}
and $T=\eta_{\mu\nu}T^{\mu\nu}$ is the scalar stress-energy.
In 3 space dimensions eq. (\ref{eulerlagrange}) is readily solved to yield
the Newtonian potential \cite{ryder}:
\begin{equation}\label{newtonianlimit}
h_{\mu\nu}(x,t)=\frac{4G^{(3)}}{{\rm c}^2}\int {\rm d}^3x\prime\frac{\tilde{T}_{\mu\nu}(x\prime ,t\prime) 
- \frac{1}{2}\eta_{\mu\nu}\tilde{T}(x\prime ,t\prime)}{\mid x-x\prime\mid}
\end{equation}
where $t\prime =t-\frac{\mid x-x\prime\mid}{c}$.
If $T_{00}=\rho c^2$ is the only non-vanishing element ($\rho$: mass density), 
one has $h_{00}=\frac{2}{{\rm c}^2}\Phi_{Newton}=\frac{2}{{\rm c}^2}\frac{G^{(3)}M_{ext}}{r}$ 
\cite{cheng2}.

\subsection{Emerging gravonons: the massive particles of a non-relativistic\\ 
gravity field}\label{gravononsect}
\noindent For a free gravitational field the Euler-Lagrange equation (\ref{eulerlagrange})
reduces to
\begin{equation}\label{kleingordonhom}
\partial^2 h_{\mu\nu} = 0
\end{equation}
As particles of molecular size couple only to short wavelength gravitons
(i.e., quantized gravitational waves) we can write the relativistic energy $E$ 
of the gravitons as
\begin{equation}
E=k{\rm c}+\epsilon
\end{equation}
with $k=\mid \vec{k}\mid $ and $\vec{k}$ labelling the Fourier component
of $h_{\mu\nu}$. 
(In this section the velocity of light ${\rm c}$ will be explicitly displayed.)
With $k$ around $10$ bohr$^{-1}$ $E$ will be of the order of $10^3$ Hartree, whereas
$\epsilon$ characterizes the energy scale with which the boson matter field $\phi$
couples to the gravitons
and is of the order of $10^{-12}$ Hartree. 
(We need short wavelength gravitons with $\lambda \approx 1$ bohr for the localization
of particles of atomic size.)
The energy $E$ and the length $k$ of the
wave vectors involved are therefore defined with a relative
precision of $10^{-15}$. This motivates us to write for the graviton field:
\begin{equation}\label{gravonondef}
h_{\mu\nu}(x)=e^{-ik{\rm c}t}\zeta_{\mu\nu}(x).
\end{equation}
{\em The gravonons} $\zeta_{\mu\nu}$ contain the gravitational potential and are, of course, not simple number operators.
Writing eq. (\ref{kleingordonhom}) (Klein-Gordon equation for the field $h_{\mu \nu}$ with $m=0$) as
\begin{equation} \label{kleingordon2}          
(\frac{\partial ^2}{\partial t^2}-{\rm c}^2\nabla ^2)h_{\mu\nu}=0
\end{equation}
and, plugging in eq. (\ref{gravonondef}),
yields the following relationships:
\begin{eqnarray}\label{firstderiv}
\frac{\partial}{\partial t}h_{\mu\nu}&=&(-ik{\rm c})h_{\mu\nu}
+e^{-ik{\rm c}t}\frac{\partial}{\partial t}\zeta_{\mu\nu}
\end{eqnarray}
The second derivative is then
\begin{eqnarray}
\frac{\partial^2}{\partial t^2}h_{\mu\nu}&=&(-ik{\rm c})e^{-ik{\rm c}t}\left [ -ik{\rm c}\zeta_{\mu\nu}
+2\frac{\partial}{\partial t}\zeta_{\mu\nu}
+\frac{1}{(-ik{\rm c})}\frac{\partial^2}{\partial t^2}\zeta_{\mu\nu} \right ].
\end{eqnarray}
With $\frac{\partial^2}{\partial t^2}\zeta_{\mu\nu}$ being of order $\epsilon^2\approx 10^{-24}$
Hartree$^2$ this term can safely be omitted and we obtain for eq. (\ref{kleingordon2}):
\begin{equation}
-ik{\rm c}\zeta_{\mu\nu}+2\frac{\partial}{\partial t}\zeta_{\mu\nu}=-\frac{\rm c}{ik}\nabla^2\zeta_{\mu\nu}.
\end{equation}
Multiplying by $\frac{i}{2}$ yields the Schr\"odinger equation:
\begin{equation}\label{schroedgravonon}
i\frac{\partial}{\partial t}\zeta_{\mu\nu}=(-\frac{\nabla^2}{2m_g}+V_o)\zeta_{\mu\nu}
\end{equation}
with $m_g=\frac{k}{\rm c}$ the mass of {\em the emerging gravonon} and $V_o=-\frac{k{\rm c}}{2}$
a constant potential cancelling the zero point energy of the light cone graviton
$h_{\mu\nu}(\vec{k})$.
With $k=10$ bohr$^{-1}$ the mass $m_g$ of the gravonon is roughly a tenth of the mass of an electron.

\subsubsection{The free gravonon action}
\noindent In this section the action of the free and the 
perturbed gravonon field via $\delta S_{matter}$ is derived.
The Schr\"odinger equation (\ref{schroedgravonon}) can be obtained by varying the 
following non-relativistic action
with respect to $\zeta_{\mu\nu}^+$:
\begin{equation} \label{lagrangegravonon}
S_{gravonon}=\int {\rm d}t L_{gravonon}=\int {\rm d}t{\rm d}^Dx\left[i\zeta_{\mu\nu}^+\frac{\partial}{\partial t}
{\zeta_{\mu\nu}}+\frac{1}{2m_g}\zeta_{\mu\nu}^+\nabla^2\zeta_{\mu\nu}
-\zeta_{\mu\nu}^+V_o\zeta_{\mu\nu}\right]
\end{equation}
in $D$ spatial dimensions.
This is, however, not the action obtained by inserting eq. (\ref{gravonondef})
in $S_{wfg}-\delta S_{matter}$ (cf. eq. \ref{weakfieldgravity}). One has:\\
\begin{equation} 
2m_gS_{gravonon}= S_{wfg}-\delta S_{matter}
\end{equation}

\noindent
Determining $\zeta_{\mu\nu}^+, \zeta_{\mu\nu}$ from eq. (\ref{lagrangegravonon})
yields results for $\sqrt{2m_g}\zeta_{\mu\nu}$. Therefore such a solution has to be
divided by $\sqrt{2m_g}$ before plugging into $T_{00}$.\\

\noindent
Potential terms can be constructed from eq. (\ref{interaction}) together with eq. (\ref{tmunulin})
yielding expressions like $\zeta_{\mu\nu}L_{matter}$, which contain terms linear and quadratic
in $\zeta_{\mu\nu}$.  The linear terms will then be taken into account by shifting the
gravonon operator whereas the quadratic terms are inserted into eq. (\ref{schroedgravonon})
before solving for $\zeta_{\mu\nu}$.

\subsection{Matter fields}\label{matterfields}
A massive scalar boson field $\phi$ representing the matter fields is introduced as \cite{zee}
\begin{equation} \label{scalarboson}
S_{matter}=-\int {\rm d}^4x\sqrt{-g}(\frac{1}{2}g^{\mu\nu}\partial_\mu \phi^+ \partial_\nu \phi
-\frac{1}{2}m^2\phi^+\phi)
\end{equation}
$\phi$ lives only in four dimensional spacetime. It is extended to
$N$ dimensional spacetime by setting all values of $\phi(x)$ having $x^ \lambda >0$ for $\lambda > 3$
to zero. \\

\noindent
The stress-energy tensor is obtained from eq. (\ref{stressenergy}) by varying the action of the matter fields
with respect to the metric: 
\begin{equation} \label{tmunu}
T_{\mu\nu}=
\partial_\mu \phi^+ \partial_\nu \phi
-g_{\mu\nu}(\frac{1}{2}g^{\rho\lambda}\partial_\rho \phi^+ \partial_\lambda \phi
-\frac{1}{2}m^2\phi^+\phi).
\end{equation}
This is a textbook result and it is valid in general before linearization.
Before plugging this into eq. (\ref{eulerlagrange}) we linearize
the stress-energy tensor by utilizing eq. (\ref{lineargrav}) in the following way:
\begin{eqnarray} \label{tmunulin}
T_{\mu\nu}&=&
\partial_\mu \phi^+ \partial_\nu \phi
-(\eta_{\mu\nu}+h_{\mu\nu})\left [ \frac{1}{2}(\eta^{\rho\lambda}+h^{\rho\lambda})\partial_\rho \phi^+ \partial_\lambda \phi
-\frac{1}{2}m^2\phi^+\phi \right ].
\end{eqnarray}
For operations on the field $\phi$ the indices run from zero to 3, for $\eta$ and $h$
they run from 0 to $N$.
In this linearized form the stress-energy tensor retains contributions from the
gravitational field $h_{\mu\nu}$ and the interaction term eq. (\ref{interaction})
of the action contains terms quadratic in the gravitational field $h_{\mu\nu}$.
This kind of self-interaction of the gravitational field turns out to be essential for 
understanding adsorbate diffusion on solid surfaces and telegraph like quantum jumps 
(cf. the analysis in ref. \cite{doyenDICE2012} and section \ref{quantumjumps}.)

\section{Non-relativistic limit of the massive scalar boson field}\label{nonrel}
\noindent The following argument is taken from ref. \cite{zee}.
The Klein-Gordon equation for a free scalar boson field of mass $m$ is:
\begin{equation} \label{kleingordon}
(\partial^2+m^2)\phi =0.
\end{equation}
For a non-relativistic particle (in the sense of special relativity) the kinetic energy is much smaller than
its mass $m$ (the velocity of light ${\rm c}$ is unity in our notation,
as well as $\hbar$)
and the relativistic energy $E$ is written as
\begin{equation} 
E=m+\varepsilon.
\end{equation}
A boson field $\psi$ varying slowly in time can then be
defined:
\begin{equation}\label{slowpsi} 
\phi (x) = e^{-imt}\psi (x).
\end{equation}
Setting 
\begin{equation}
\frac{\partial^2}{\partial t^2}\psi =0,
\end{equation}
Schr\"odinger's equation is obtained:
\begin{equation}\label{schroedmass}  
i\frac{\partial}{\partial t}\psi = -\frac{\nabla^2}{2m}\psi
\end{equation}
with $\nabla = (\frac{\partial}{\partial x^1},\frac{\partial}{\partial x^2}, ...,\frac{\partial}{\partial x^D})$
for $D$ spatial dimensions. For more details see ref. \cite{zee}.\\

\noindent
As the matter field $\psi$ lives only in  four dimensional spacetime,
the Schr\"odinger equation can be obtained by varying the following non-relativistic action
with respect to $\psi^+$:
\begin{eqnarray} \label{lagrangeschroed}
S_{matter-nonrel}&=&\int {\rm d}t{\rm d}^3x{\cal L}_{matter-nonrel}=\int {\rm d}t L_{matter-nonrel}\nonumber \\
&=&\int {\rm d}t{\rm d}^3x\left[i\psi^+\frac{\partial}{\partial t}\psi+\frac{1}{2m}\psi^+\nabla^2\psi\right].
\end{eqnarray}
This form of the action linear in $\frac{\partial}{\partial t}$ is obtained by partial integration
(cf. ref. \cite{zee}).
This is, however, not the action obtained by inserting eq. (\ref{slowpsi})
in $S_{matter}(h_{\mu\nu}=0)$ (eq. \ref{scalarboson}). One has:
\begin{equation} 
2mS_{matter-nonrel}= S_{matter}(h_{\mu\nu}=0).
\end{equation}
The variation of the action eq. (\ref{lagrangeschroed}) yields a solution for $\sqrt{2m}\psi$. The solution
obtained in this way must hence be divided by $\sqrt{2m}$ before
plugging into $T_{00}$.

\subsection{The matter - gravonon coupling}
\noindent The Lagrangian of the matter-gravonon interaction is the focus in the present
section.
In the nonrelativistic limit only $h_{00}$ and $T_{00}$ need to be considered, all other
components being smaller by a factor $(\frac{v}{\rm c})^2$ where $v$ is the nonrelativistic
velocity. Plugging eqs. (\ref{slowpsi}) and (\ref{gravonondef}) into eq. (\ref{tmunulin}) yields
($\eta_{00}=1$):
\begin{eqnarray} \label{tmunu00}
T_{00}&=&
\partial_0 \phi^+ \partial_0 \phi-(\frac{1}{2}\partial_0 \phi^+ \partial_0 \phi-\frac{1}{2}m^2\phi^+\phi)\nonumber \\
&-&h_{00}(\frac{1}{2}\partial_0 \phi^+ \partial_0 \phi-\frac{1}{2}m^2\phi^+\phi)
-\frac{1}{2}h^{00}\partial_0 \phi^+ \partial_0 \phi\nonumber \\
&=&\partial_0 \phi^+ \partial_0 \phi-(\frac{1}{2}\partial_0 \phi^+ \partial_0 \phi-\frac{1}{2}m^2\phi^+\phi)
+h_{00}\frac{1}{2}m^2\phi^+\phi\nonumber \\
&=&\frac{1}{2}\partial_0 \phi^+ \partial_0 \phi+\frac{1}{2}m^2\phi^+\phi
+h_{00}\frac{1}{2}m^2\phi^+\phi\nonumber \\
\end{eqnarray}
(We use $h_{00}=-h^{00}$; $\psi \partial_0 \psi^+=\psi^+ \partial_0 \psi$;
$\phi(x)=e^{-imt}\psi(x)$; $\phi^+(x)=e^{imt}\psi^+(x)$; $h_{00}=e^{-ik{\rm c}t}\zeta_{00}$.)
Lines 1 and 2 are the standard textbook result. Line 3 contains terms arising from
$h_{\mu\nu}$ in eq. (\ref{tmunulin}).
From line 3 to line 4 we used $h^{00}=-h_{00}$ which follows from
\begin{eqnarray}
(1+h_{00})^{-1} &=& 1+ h^{00}\nonumber\\
&\approx &1-h_{00}.
\end{eqnarray}
Up to here the expression is relatistically invariant. We now insert the definition eq. (\ref{slowpsi}).
\begin{eqnarray} \label{tmunu00-2}
T_{00}&=&\frac{1}{2}[m^2\psi^+\psi+\partial_0 \psi^+ \partial_0 \psi+im \psi^+ \partial_0 \psi-im \psi \partial_0 \psi^+]\nonumber \\
&+&\frac{1}{2}m^2\phi^+\phi+\frac{1}{2}e^{-ik{\rm c}t}\zeta_{00}m^2\phi^+\phi\nonumber \\
&=&m^2\psi^+\psi+\frac{1}{2}\partial_0 \psi^+ \partial_0 \psi
+\frac{1}{2}e^{-ik{\rm c}t}\zeta_{00}m^2\psi^+\psi\nonumber \\
&\approx&m^2\psi^+\psi(1+\frac{1}{2}e^{-ik{\rm c}t}\zeta_{00}).
\end{eqnarray}
In the first two lines of eq. (\ref{tmunu00-2}) we used eq. (\ref{gravonondef}). 
From lines 1 and 2 to line 3 we used the ansatz eq. (\ref{slowpsi}).
In the last line we used that $\psi$ is slowly varying in time and
hence $\partial_0 \psi^+ \partial_0 \psi$ is much smaller than $\frac{1}{2}m^2\psi^+\psi$.\\

\noindent
Now in the lowest order, the Newtonian limit, the expectation value of $T_{00}$ has to be
\begin{equation}
\langle \Psi\mid T_{00}\mid\Psi\rangle =\rho {\rm c}^2 =m{\rm c}^2\langle \psi^+\psi\rangle
\end{equation}
with $\mid \Psi \rangle$ the world wave function. However, using eq. (\ref{tmunu00-2}),
\begin{equation}
\langle \Psi \mid T_{00} \mid\Psi\rangle=\langle \Psi \mid m^2 \psi^+\psi \mid \Psi\rangle=
m^2 \langle \Psi\mid \psi^+\psi \mid \Psi\rangle.  \nonumber
\end{equation}
This can only be achieved with\\
\begin{equation}
\psi\rightarrow\sqrt{\frac{1}{m}}\psi.
\end{equation}

\noindent
If $\psi$ in eq. (\ref{tmunu00-2}) is obtained from eq. (\ref{lagrangeschroed}), then $\psi$ is too
large by a factor of $\sqrt{2m}$ and eq. (\ref{tmunu00-2}) has to be divided by $2m$.
If $\zeta_{00}$ in eq. (\ref{tmunu00-2}) is obtained from eq. (\ref{lagrangegravonon}), 
then $\zeta_{00}$ is too
large by a factor of $\sqrt{2m_g}$ and $\zeta_{00}^+\zeta_{00}$ 
in eq. (\ref{tmunu00-2}) has to be divided by $2m_g$.\\

\noindent
The interaction Lagrangian (matter field with gravitational field), i.e., the integrand in
eq. (\ref{interaction}), using eq. (\ref{tmunu00-2}), is then:
\begin{eqnarray} \label{l00}
\frac{1}{2}h_{00}T^{00}(1-\frac{1}{2}h^{00})&=&\frac{1}{2}h^{00}T_{00}(1-\frac{1}{2}h^{00})\nonumber\\
&=&\frac{1}{2}e^{ik{\rm c}t}\zeta_{00}^+T_{00}(1-\frac{1}{2}h^{00})\nonumber \\
&\approx&\frac{1}{2}m^2\psi^+\psi(e^{ik{\rm c}t}\zeta_{00}^++\frac{1}{2}\zeta_{00}^+\zeta_{00})(1-\frac{1}{2}h^{00})\nonumber\\
&=&\frac{1}{2}m^2\psi^+\psi(h_{00}+\frac{1}{2}\zeta_{00}^+\zeta_{00}-\frac{1}{2}h_{00}h^{00}
-\frac{1}{4}\zeta_{00}^+\zeta_{00}h^{00})
\end{eqnarray}
where eq. (\ref{gravonondef}) has been used in the second line, $h_{00}$ is real:
\begin{equation} \label{h00-h00star}
h_{00}=h_{00}^*=-h^{00}=-h^{00*}=e^{-ikct}\zeta_{00}=e^{ik{\rm c}t}\zeta_{00}^+
\end{equation}
implying $h_{00}T^{00}=h^{00}T_{00}$ and $h_{00}h^{00}=\zeta_{00}^+\zeta_{00}$.
In 4-dim spacetime
$h_{00}$ is proportional to the Newtonian potential obtained from eq. (\ref{newtonianlimit})
and according to ref. \cite{cheng2} should be equal to 
$h_{00}=\frac{2}{{\rm c}^2}\Phi_{Newton}=\frac{2}{{\rm c}^2}\frac{G^{(3)}M_{ext}}{r}$.
Observe that there is no term linear in $\zeta_{00}$ in eq. (\ref{l00}), i.e., there is no
dipole interaction between the matter field and the gravonons. The terms involving
$\zeta_{00}^+\zeta_{00}$ (arising from terms $h^{00}h^{00}$ and $\zeta_{00}^+h^{00}$)
signify gravonon-gravonon interaction. They contain the square of the gravitational potential.\\

\noindent
The Lagrangian describing the interaction of the scalar boson with Newton's potential
and the gravonon field is then (omitting the higher order term proportional to $h_{00}h^{00}$
and setting 
$h_{00}=\frac{2}{{\rm c}^2}\frac{G^{(D)}M_{ext}}{r^{D-2}}$ and c=1):
\begin{eqnarray}\label{lint}
L_{int} &=& \int {\rm d}^Dx {\cal L}_{int}\nonumber\\
&=&\int {\rm d}^Dx \left[\frac{G^{(D)}mM_{ext}}{r^{D-2}}\psi^+\psi+\frac{m}{2}\zeta^+_{00}\zeta_{00}\psi^+\psi
-\frac{1}{4}\frac{G^{(D)}mM_{ext}}{r^{D-2}}\zeta^+_{00}\zeta_{00}\psi^+\psi \right].
\end{eqnarray}
$D$ is the number of spatial dimensions.
The first term in the second line might be termed the Schr\"odinger-Newton contribution as 
in four dimensional spacetime it is
the interaction term in the Schr\"odinger-Newton equation \cite{penrose1}.
The second term is the scalar boson - gravonon interaction. The third term
describes the interaction between the Schr\"odinger-Newton term and the gravonons
and renormalizes Schr\"odinger-Newton term:
\begin{eqnarray}\label{lintrenorm}
L_{int} &=& \int {\rm d}^Dx \left[\frac{G^{(D)}mM_{ext}}{r^{D-2}}(1-\frac{1}{4}\zeta^+_{00}\zeta_{00})\psi^+\psi
+\frac{m}{2}\zeta^+_{00}\zeta_{00}\psi^+\psi \right].
\end{eqnarray}

\subsection{Effective Schr\"odinger equations}
\noindent Adding ${\cal L}_{int}$ (eq. \ref{lintrenorm}) to the Lagrangian density
${\cal L}_{matter-nonrel}$ of eq. (\ref{lagrangeschroed})
and varying with respect to $\psi^+$ yields the Schr\"odinger equation
\begin{equation}
i\frac{\partial}{\partial t}\psi = \left[-\frac{\nabla^2}{2m}
-\frac{G^{(D)}mM_{ext}}{r^{D-2}}(1-\frac{1}{4}\zeta^+_{00}\zeta_{00})
-\frac{m}{2}\zeta^+_{00}\zeta_{00}\right]\psi
\end{equation}
which we have to solve simultaneously with the Schr\"odinger equation
arising from varying ${\cal L}_{gravonon}$ (eq. \ref{lagrangegravonon}) and 
${\cal L}_{int}$ of eq. (\ref{lintrenorm}) with respect to $\zeta^+_{00}$
\begin{equation}
i\frac{\partial}{\partial t}\zeta_{00} = \left[-\frac{\nabla^2}{2m_g}
-\frac{m}{2}\psi^+\psi
+\frac{1}{4}\frac{G^{(D)}mM_{ext}}{r^{D-2}}\psi^+\psi\right]\zeta_{00}+(V_o-\frac{kc}{2}h_{00})\zeta_{00}.
\end{equation}

\subsection{Effective Lagrangian}
\noindent The effective Lagrangian is now constructed by adding the Lagrangians contained
in the actions of eqs. (\ref{lagrangeschroed}), (\ref{lagrangegravonon}) and (\ref{lintrenorm})
\begin{equation}
L_{eff}=L_{matter-nonrel}+L_{gravonon}+L_{int}
\end{equation}
From this
the effective Hamiltonian is deduced
with the help of the familiar relationship
\begin{equation}
H_{eff}=\int {\rm d}^Dx\frac{\partial L_{matter-nonrel}}{\partial (\partial_t\psi)}\partial_t\psi
+\int {\rm d}^Dx\frac{\partial L_{gravonon}}{\partial (\partial_t\zeta_{00})}\partial_t\zeta_{00}
-L_{eff},
\end{equation}
yielding:
\begin{eqnarray} \label{effectivehamilt}
H_{eff}&=&H^o_{gravonon}-L_{int} \nonumber \\
&=&\int {\rm d}^Dx\left[-\psi^+\frac{\nabla^2}{2m}\psi -\zeta^+_{00}\frac{\nabla^2}{2m_g}\zeta_{00}
+(V_o-\frac{kc}{2}h_{00})\zeta^+_{00}\zeta_{00}\right] -L_{int}.
\end{eqnarray}
This is the quantum field theoretical Hamiltonian which we have to solve,
as it is described in section \ref{charsolution}, for the problems of quantum particles localization, 
the diffusion problem, wave-to-particle transition, etc..

\section{Model for the non-interacting and the matter-induced \\
 gravonon structure}\label{gravstruc}
\noindent The gravonon field in the second and third term of the effective Hamiltonian
eq. (\ref{effectivehamilt}) is a free field in a constant potential. The adsorbate
motion, which we want to investigate, occurs, however, in the presence of other  
masses, which have already induced a structure in the gravonon field.
The modelling of the pre-existing gravonon structure has been described in 
a previous paper \cite{drakovaDICE2012}. \\ 

\noindent The modification of the gravonon structure induced by matter fields has 
also been described in ref. \cite{drakovaDICE2012} and refers to 
the generation of soft gravonons.
The soft gravonon mode is an adsorbate
induced coherent motion within a spacetime deformation involving 5 to 10 atoms 
in the solid surface. 
The local intrinsic gravonon coherence is retained unless
it is distroyed by other fields in the environment.\\

\noindent
In eq. (\ref{effectivehamilt}) we develop the gravonon fields as follows:
\begin{eqnarray}\label{gravmodexp}
\zeta_{00}&=&\sum_ib_i\zeta_{00}^{(i)}V_{grav}(x_i)\theta  \nonumber \\
\zeta_{00}^+&=&\sum_ib_i^+\zeta_{00}^{*(i)}V_{grav}(x_i)\theta. 
\end{eqnarray}
$\zeta_{00}^{(i)}$, $\zeta_{00}^{*(i)}$ are functions localized near the positions $x_i$
of the atomic cores.
The gravitational potential $V_{grav}(x_i)$ has been extracted from the localized
fields. The factor $\theta$ serves to make the expansion coefficients $b_i$, $b_i^+$
dimensionless.   
The effective Hamiltonian
eq. (\ref{effectivehamilt}) provides for the term $H^o_{gravonon}$:
\begin{eqnarray}
H^ o_{gravonon}&=&\int {\rm d}^Dx\zeta^+_{00}(-\frac{\nabla^2}{2m_g}+V_o)\zeta_{00}\nonumber\\
&=&\sum_{i,j}\langle\zeta^{*(i)}_{00}\mid V_{grav}(x_i)(-\frac{\nabla^2}{2m_g}+V_o)V_{grav}(x_j)\mid \zeta_{00}^{(j)}\rangle\theta^ 2b_i^+b_j\\
&=&\sum_{i,j}\Omega_{ij}b_i^+b_j
\end{eqnarray}
where $\langle ... \rangle$ indicates the volume integration in the $D$-dimensional space and $\Omega_{ij}$
is defined by comparing the last two lines.
This is a mode expansion as it is common in quantum field theory. $b_i,b_j^+$ are boson operators
obeying the commutation relations $[b_i,b_j^+]=\delta_{ij}$. $\theta$ is such that $\Omega_{ij}$ has dimension of energy.\\

\noindent
If we diagonalize at this place, we obtain a collection of independent harmonic oscillators:
\begin{equation}\label{gravonondiag}
H^o_{gravonon}=\sum_k\Omega_k b_k^+b_k .
\end{equation}
The $\Omega_k$ have to be modelled according to the physical situation under investigation. 
For this purpose we write the boson operators $b_k$, $b_k^+$ in terms of generalized coordinates
and momenta:
\begin{eqnarray} \label{boper}
b_k^+&=&(\frac{-i}{\sqrt{2\omega_k}}p_k+\sqrt{\frac{\omega_k}{2}}q_k) \nonumber \\
b_k&=&(\frac{i}{\sqrt{2\omega_k}}p_k+\sqrt{\frac{\omega_k}{2}}q_k).
\end{eqnarray}
The diagonal form eq. (\ref{gravonondiag}) is in the present context considered
to represent the eigenmodes of oscillating spacetime deformations centered around atomic cores coupled
to each other by terms quadratic in the displacements of the cores. In this case we have from
eqs. (\ref{gravmodexp}) and (\ref{boper}):
\begin{equation}
\zeta_{00}^+(x)+ \zeta_{00}(x) = \sum_i 2q_ig(x-x_i)
\end{equation}
with 
\begin{equation}
g(x-x_i) = \sqrt{\frac{\omega_i}{2}} \zeta_{00}^{i}V_{grav}(x_i)\theta
\end{equation}
where we now assume that $\zeta_{00}^{i}$ is real.
These last two equations should replace eqs. (12) and (13) in reference \cite{drakovaDICE2012}.
For the ``potential term`` in $\zeta_{00}^+(x)\zeta_{00}(x)$ we have then from eqs. (\ref{gravmodexp})
and (\ref{boper}):
\begin{equation}\label{potharm}
\zeta_{00}^+(x)\zeta_{00}(x)=...+\sum_{i,j}q_iq_jg(x-x_i)g(x-x_j).
\end{equation}
This equation should replace eq. (14) in reference \cite{drakovaDICE2012}.
In order to describe effects due to mass motion one has to add terms to eq. (\ref{potharm})
which are not solely a function of $x-x_i$ (many-body forces). This has been described
in reference \cite{drakovaDICE2012}.

\subsection{Induced scattering in the gravonon continuum leads to quantum jumps}\label{quantumjumps}
\noindent 
The equations derived in sections \ref{nonrel} and \ref{gravstruc} have been applied
to problems of adsorbate localization \cite{doyenDICE2012}, adsorbate diffusion \cite{drakovaDICE2012}, 
scattering of massive particles from nano lattices \cite{unpublished}, behaviour of quantum computers
\cite{unpublished}. In the case of adsorbate diffusion a discrepancy of nine orders of magnitude between
the results of Copenhagen quantum mechanics 
and experiment could be removed. Our theory allows to determine the adsorbate - surface
distances from the experimental diffusion rates which are then found to be in good
agreemant with distances determined from ab-initio density functional calculations \cite{drakovaDICE2012,unpublished}.
In these model studies
quantum jumps arise as sudden changes of the weights of two or more configurations
with time. In Copenhagen quantum mechanics 
collapses of the wave function are postulated to result in quantum jumps, whereas in the present 
Emerging quantum mechanics they appear
as solutions of Sch\"odinger's equation within a unitary coherent quantum description \cite{arxive}.
In the limit of {\it a very local and weak interaction with the gravonons}   
the dynamics of an adsorbed atom between two adsorption sites $\alpha$ and 
$\beta$ on a solid surface was described by the time development of  
the occupation of the sites.
The occupation of each site was expressed as the sum of the 
weights of those configurations which involve site
$\alpha$ or $\beta$ site, respectively. 
The time variation of these configurations exhibit the telegraph character and warrant that the occupation of 
each site jumps between two values (cf. fig. \ref{telegraph-jumps}).

\section{Interpretation and ontology of Emerging QM} \label{ontologyEQM}

\noindent Emerging QM emerges from a quantum field theory which is Lorentz invariant:
\begin{itemize}
\item The world wave functional is a result of the entanglement between
local massive particles and the gravonon field in hidden spatial dimensions. 
The wave functional in Emerging QM does not have the meaning of a probability amplitude. 
No density matrix and tracing out are needed or involved.
\item Massive non-relativistic locally modified soft modes named gravonons are generated
in all spacetime dimensions, including the hidden dimensions, due to local gravitational
interactions. They modify the time development
of the local system in the real 3+1 dimensional world. Beables constitute the connection to experience.
\item Observables are not needed since measurement is treated as a quantum process.
\item Quantum jumps are intrinsic to the wave functional of Emerging QM and they 
result from the solution of the time dependent Schr\"odinger eqution in high dimensional spacetime.
They are due to the entanglement of the local system to gravonons  
of high density of states which propagate
in hidden spatial dimensions.
\item How do we come to the real 3+1 dimensional world? 
In all experiments discussed in the introduction we see {\em effects which occur locally}:  
either probe particles in the experimental source (electrons, photons, adsorbed particles) or  
electrons and photons on oscillograph screens and photographic plates,   
adsorbed particles localized on definite
adsorption sites, etc.. We do not see these particles in their flight from the source to the
target. And if we would try to see them (to ''measure'' them) on their way to the target
the result of the final measurement would be different.  
Hence we may conclude that what we see occurs locally and then
it is in the 3+1 dimensional world. The transition  
from the high-dimension configuration space to the 3+1 dimensional real world
in Emerging QM occurs because of the localization of particles via entanglement  
to gravonons. Just those local components of the wave functional
defined in high-dimensional configuration space, which are entangled with the gravonons,
define the beables and 
represent real particles in the 3+1 real world.

\item ''Measurement'' in Emerging QM is the occurrence of entanglement of a local system
with the gravonons. 
Different outcomes are the result of different initial states.
Apparent statistics in the outcome is the consequence of different initial states and hence different experimental conditions.
\item Nonlocality is in the high-dimensional wave functional defined in configuration space.
\item Emerging QM is Lorentz invariant. 
\item Emerging QM explains: telegraph signals (quantum jumps), low temperature adsorbate diffusion, 
adsorbate desorption induced by the current in the scanning tunnelling microscope, 
Penning ionization at surfaces, etc. (cf. Introduction).
\end{itemize}

\section{Comparison between Emerging QM and Copenhagen and Bohmian QM} \label{comparisonCQMEQM}
\noindent The major ontological features of Emerging QM are compared with those of 
Copenhagen and Bohmian QM \cite{duerrbook,bohm}
in table \ref{ontology}. The key double-slit experiment can be used 
to illustrate the points of view in the three theories. \\ 

\begin{itemize}
\item Copenhagen QM: Wave-particle duality, 
collapse and probability interpretation are used in a phenomenological way to provide  
the explanation of how the diffraction pattern is created on the  
detection screen. 
Schr\"odinger's wave function interferes with itself. At different screen positions 
the ''measurement'' consists of instantaneous transformation of   
waves into particles (called collapses) with 
probabilities, determined by Born's rule, namely by the modulus squared of the amplitude of 
the wave function at that position. The interference pattern in the wave
function is imprinted on the detection screen via probabilistic collapses of the wave 
function at different positions on the screen at different times. 

\item Bohm's theory: The wave function due to standard quantum mechanics  
develops the interference pattern when it interferes with itself.
Single particle trajectories pass through one slit only, however the interference  
pattern, which develops in the wave function, 
guides the particles in regions of constructive interference, avoiding    
regions of destructive interference. 
The final position of a particle on the screen and 
which slit it will choose to pass through is determined by the initial position 
of the particle (which is not controlled by the experimenter) and by the time
development of the wave function.
In the 3+1 real world this is expressed by the existence of 
regions where particle trajectories are dense and regions
with less dense particle trajectories, providing the diffraction pattern on the   
detection screen.

\item Emerging QM describes the double-slit experimental result as due to the world wave functional 
propagating through both slits and interfering with itself. The    
interference pattern is built into the wave functional already in the  
vacuum region between the source, the slits and the detection screen. 
On the screen, where matter density is high, components of the 
world wave functional  
entangle locally with the gravonons, leading to particles localizing on the screen.  
This occurs more often with components,  
which have high amplitude in the total wave functional, i.e. 
particle localization at selected positions on the screen occur more often for repeated experiments,     
giving rise to the diffraction pattern. \\
\end{itemize}

\noindent In table \ref{ontology} an attempt is made to identify similarities  
and differences in the ontologies of Copenhagen quantum mechanics, Bohm's QM and Emerging QM.
\begin{center}
\begin{table}{}  
\begin{minipage}{15cm}
\caption{\label{ontology}Comparison between the ontology within Copenhagen QM, Bohm's QM
and Emerging QM}
\begin{tabular}{|l|c|c|c|} 
\hline
&&&\\
{\bf ontological property}      &{\bf Copenhagen QM} & {\bf Bohm's QM}      &{\bf Emerging QM}\\
&&&\\
\hline
determinism                     & NO                 &YES                   & YES\\
\hline
chooser                         & NO (``God plays dice``)&YES (particle positions)& YES (gravonons)\\
\hline
collapse                        & YES                &NO                    & NO \\
\hline
measurement problem             & YES                &NO                    & NO \\
\hline
probabilistic                   & YES                & NO                   & NO \\              
interpretation                  &                    &                      &  \\
\hline
statistics ditribution          & NO                 & YES                  & YES    \\
in initial state                &                    &                      &        \\
\hline
hidden physical                 & NO                 & YES                  & YES \\
quantities                      &                    & (positions)          & (gravitons in hidden\\
                                &                    &                      & dimensions)\\
\hline
definite outcome                & NO                 & YES                  & YES  \\
\hline
pre-existing realism in         & NO                 & YES                  & YES \\              
3+1 dimensions                  &                    &                      &  \\
\hline
particles in                    & NO                 & YES                  & YES \\              
3+1 dimensions                  &                    &                      &  \\
\hline
\hline
quantum                         & NO                 & YES                  & NO \\              
equilibrium                     &                    &                      &  \\
\hline
wave function is                & NO, needs          & NO                   & YES    \\
complete description            & probabilistic Interpret.&              &  \\
\hline
\hline
Lorentz invariance              & NO                 & NO                   & YES    \\
                                &                    &               & (it is the limit of a \\
                                &                    &               & Lorentz invariant theory)\\
\hline
new results compared                &                   & NO                  & YES    \\
to Copenhagen QM                &                    &                     &  \\
\hline
\end{tabular}
\end{minipage}
\end{table}
\end{center}

\noindent As it is evident from table \ref{ontology} there are many features 
shared by Bohmian QM and Emerging QM. 
We emphasize two point: (i) Both for Bohmian QM and Emerging QM 
the measurement problem and collapse do not exist and  
(ii) Particles in 3+1 dimensional space exist, i.e. local realism is implicit.     
In Emerging QM these are the configurations which entangle with the gravonons and where the
particles are localized at definite positions at definite time. 
In Bohmian QM these are particles on definite trajectories. \\

\noindent An important difference is that Bohmian QM is not,
whereas Emerging QM is Lorentz invariant, being the non-relativistic limit 
of Quantum field theory. Superluminal changes    
of particle position are possible in Emerging QM in the absolute non-relativistic  
limit of extremely weak interactions between the local system  
and environmental gravonons of zero frequency  
which eventually might violate causality. However, in all practical
situations a {\em large but finite velocity} in the   
behaviour of the local system is obtained in the 3+1 world as solution of  
the time dependent Schr\"odinger equation.\\

\noindent Emerging QM provides the description of many other experiments which
is not possible for Bohmian QM. To mention some of them: 
\begin{itemize}
\item Adsorbate localization and slow down of diffusion on solid surfaces.
\item Telegraph like quantum jumps of adsorbates between adsorption sites in the regime
of quantum diffusion at low temperatire, resulting from the time dependent 
Schr\"odinger equation.
\end{itemize}

\section{Conclusion}

\noindent Emerging QM, the focus of the present article, is derived from  
Quantum field theory in the non-relativistic limit. It complies therefore 
with Lorentz invariance and does not suffer from this deficit in Copenhagen QM and   
Bohmian QM. Quantum jumps in Emerging QM do not violate causality.\\

\noindent The theory accounts for nonlocal correlations as a result of local entanglement    
of particles with the high-dimensional gravitational field.
Weak and local interaction with the gravitational filed of high density of states 
leads to particle localization and local realism.
Starting from high-dimensional configuration space, 
we can go to the 3+1 spacetime as a result of particle localization  
via entanglement with the gravitational field and explain experimental observations
in the real 3+1 dimensional world, not violating Lorentz invariance.\\ 

\noindent Emerging QM is realistic in a sense similar to Bohm's quantum mechanics. 
At the same time it is capable of reproducing and explaining  
experimental results which are outside the scope of Copenhagen QM   
and Bohmian QM. Within Emerging QM particle localization,
adsorbate diffusion on solid surfaces and quantum jumps, as they are 
experimentally measured in the real 3+1 world,
are the result of the solution of Schr\"odinger's 
time dependent equation for a local system which  
entangles with the high density of states of the gravitational field. \\

\noindent A message of the present contribution is that there is no probabilistic
element in the time development resulting from the time dependent   
Schr\"odinger equation. The time development
of a local quantum system entangled to quantum fields in its environment of
high dimensions and high density of states, though it 
resembles telegraph signal-like quantum jumps, is deterministic.
The conditions for the quantum jumps are: weak and local interaction with the environmental
degrees of freedom (the gravitational field) with high density of states.
Statistical appearance of experiments arises due to different 
initial states which are not controlled in experiment.\\

\noindent 
The connection from the   
high-dimensional configuration space 
to the real 3+1 world 
arises because coupling to the gravitational field 
occurs only for matter fields localized in three dimensional space
and this reproduces the results of experiments in the real world
without the need of collapse and state reduction, just as a result of the 
time development of a local system entangled with the gravitational field continuum, 
as Schr\"odinger's equation requires. In the extreme non-relativistic limit of vanishingly weak     
interaction with gravitational field modes of zero frequency the theory  
may lead to superlumical quantum jumps. 
But in all practical situations this is not the case.  
Being derived from quantum field theory, Emerging QM is Lorentz invariant 
and is not in conflict with the concept of causality. \\

\noindent The derivation of the equations of Emerging QM and their application to     
reproduce and understand concrete experiments leads to the conclusion that 
Schr\"odinger quantum mechanics has emerged. Furthermore Emerging QM  
allows the interpretation of quantum mechanics in the real 3+1 world without
the concepts of collapse and Born's probability. The crucial feature of Emerging QM is 
the localization of quantum particles     
due to entanglement with the modes of the gravitational field in high-dimensional spacetime.\\

%\noindent Declaration: The authors declare that they have no conflict of interest. 


\begin{thebibliography}{999}

\bibitem{penrose1} Penrose R.: Gen. Rel. Grav. {\bf 28}, 581 (1996)
\bibitem{penrose2} Penrose R.: Phil. Trans. R. Soc. A {\bf 356}, 1927 (1998)
\bibitem{penrose3} Penrose R.: The Large, the Small and the Human Mind. Cambridge 
Univ. Press, U.K. (1997)
\bibitem{penrose4} Penrose R.: The Emperor's New Mind. Oxford Univ. Press, Oxford (1992)
\bibitem{penrose5} Penrose R.: Shadow of the Mind. Oxford Univ. Press, Oxford (1994)
\bibitem{penrose6} Penrose R.: The Road to Reality - A Complete Guide to the Laws  
of the Universe. Knopf, New York (2005)
\bibitem{omnes} Omn\a`es R.: Are there unresolved problems in the interpretation of
quantum mechanics? In: Breuer H.-P., Petruccione F. (eds.) Open Systems and Measurement 
in Relativistic Quantum Theory. Springer, Berlin Heidelberg New York (1999) 
\bibitem{kiefer1} Kiefer C.: Quantum Gravity. Oxford Univ. Press, Oxford (2007)
\bibitem{kiefer2} Kiefer C.: On the interpretation of quantum theory - from Copenhagen 
to the present day. In: Castell L., Ischebeck O. (eds.) Time, Quantum and Information, pp 291-292.
Springer, Berlin (2003)
\bibitem{kieferjoos} Kiefer C., Joos E.: Decoherence: concepts and exmaples. In: Blanchard Ph., Jadczyk A. (eds.)
Quantum Future - From Volta and Como to the Present and Beyond.
Springer, Berlin (1999)  
\bibitem{despagnat1} d'Espagnat B.: Conceptual Foundations of Quantum Mechanics. Benjamin, 
Menlo Park Cal. (1971)  
\bibitem{despagnat2} d'Espagnat B.: Problems in objectizing. In: Blanchard Ph., Jadczyk A. (eds.)
Quantum Future - From Volta and Como to the Present and Beyond.
Springer, Berlin (1999) 
\bibitem{duerrbook} D\"urr D., Teufel S.: Bohmian Mechanics - The Physics and Mathematics of 
Quantum Theory. Springer, Berlin Heidelberg (2009)
\bibitem{vonNeumann} von Neumann J.: Mathematical Foundations of Quantum Mechanics. 
Princeton University Press, Princeton (1932/1955)
\bibitem{string} Becker K., Becker M., Schwarz J.H.: String Theory
and M-Theory.
Cambridge Univ. Press (2007); Susskind L.: The Cosmic Landscape - String
Theory and the Illusion of Intelligent Design. Back Bay Books (2006)
\bibitem{thooft1} 't Hooft G.: Class. Quant. Grav. {\bf 16}, 3263 (1999)
\bibitem{thooft2} 't Hooft G.: http://arxiv.org/pdf/quant-ph/0212095 (2002) 
\bibitem{thooft3} 't Hooft G.: Int. J. Theor. Phys. {\bf 42}, 355-361 (2003)
\bibitem{thooft4} 't Hooft G.: J. Phys.: Conf. Ser. {\bf 67}, 012015 (2007)
\bibitem{blasone} Blasone M., Jizba  P., Vitiello G.: Phys. Lett. A
{\bf 287}, 205-210 (2001)
\bibitem{adler} Adler S.L.: Quantum Theory as a Emergent Phenomenon -
The Statistical Mechanics of Matrix Models as the Precursor of Quantum
Field Theory. Cambridge Univ. Press (2004)
\bibitem{adler1} Adler S.L.: J. Phys.: Conf. Ser. {\bf 361}, 012002 (2012)
\bibitem{smolin} Smolin L.: http://arxiv.org/pdf/hep-th/0201031 (2002)
\bibitem{biro} Bir\'o T.S., Matinyan S.G., M\"uller B.:, Found. Phys.
Lett. {\bf 14}, 471 (2001)
\bibitem{hadley} Hadley M.J.: Found. Phys. Lett. {\bf 10}, 43 (1997)
\bibitem{GRW} Ghirardi G.C., Rimini A., Weber T.: Phys. Rev. D {\bf 34}, 470 (1986)
\bibitem{Pearlecoll} Pearle P.: Collapse Models. In: Breuer H.-P., Petruccione F. (eds.) 
Open Systems and Measurement in Relativistic Quantum Theory, 
p. 195. Springer (1999)
\bibitem{haag} Haag R.: Objects, Events and Localization. In: Blanchard Ph., Jadczyk A. (eds.) 
Quantum Future - From Volta and Como to the Present and Beyond,
p. 58-79, Springer, Berlin (1999) 
\bibitem{eiglerXe} Eigler D.M., Schweizer E.K.: Nature (London) {\bf 344}, 524 (1990)
\bibitem{lauhonho} Lauhon L.J., Ho W.: Phys. Rev. Lett. {\bf 85}, 4566 (2000);
Erratum: Lauhon L.J., Ho W.: Phys. Rev. Lett. {\bf 89}, 079901 (2002)
\bibitem{meyer} Liljeroth P., Repp J., Meyer G. Science {\bf 317}, 1203 (2007)
\bibitem{gomer} Lin T.S., Gomer R.: Surface Sci. {\bf 255}, 41 (1991)
\bibitem{wahn1997} Mattsson Th., Wahnstr\"om G.: Phys. Rev. B
{\bf 56}, 14944 (1997)
\bibitem{COdesorption} Bartels L., Meyer G., Rieder K.-H., Velic D., Knoesel E.,
Hotzel A., Wolf M., Ertl G.: Phys. Rev. Lett. {\bf 80}, 2004 (1998)
\bibitem{doydrakCOdesorp} Drakova D., Doyen G.: J. Phys.: Conf. Ser. {\bf 61}, 262 (2007)
\bibitem{ertl} Conrad H., Ertl G., K\"uppers J., Wang S.W., G\'erard K.,
Haberland H.: Phys. Rev. Lett. {\bf 42}, 1082 (1979);
Conrad H., Ertl G., K\"uppers J., Sesselmann W., Haberland H.:
Surface Sci. {\bf 121}, 161 (1982)
\bibitem{doyHe-COscattering} Conrad H., Doyen G., Ertl G., K\"uppers J.,
Sesselmann W.: Chem. Phys. Lett. {\bf 88}, 281 (1982)
\bibitem{doyphyrevb} Doyen G.: Phys. Rev. B {\bf 22}, 497 (1980)
\bibitem{aepli} Brooke J., Bitko D., Rosenbaum T.F., Aepli G.: Science {\bf 284}, 779 (1999)
\bibitem{dwavesys} http://www.dwavesys.com
\bibitem{chiorescu} Chiorescu I., Nakamura Y., Harmans C.J.P.M, Mooij J.E.: Science {\bf 299}, 1869 (2003)
\bibitem{johnson} Johnson M.W. et al.,  Nature {\bf 473}, 194 (2011)
\bibitem{dwave} Dickson N.G. et al., Nature Commun. 4:1903 DOI:10.1038/ncomms2920 (2013)
\bibitem{unpublished} Drakova D. , Doyen G., in preparation.
\bibitem{diosi3} Di\a'osi L.: Phys. Rev. A {\bf 40}, 1165 (1989); http://arxiv.org/quant-ph/060711 (2005)
\bibitem{penrose} Moroz I.M., Penrose R., Tod K.P.: Class. Quant. Grav. {\bf 15}, 2733 (1998)
\bibitem{ghirardi} Ghirardi G.C., Grassi  R., Rimini A.: Phys. Rev. A {\bf 42}, 1057 (1990) 
\bibitem{percival} Percival I.C.: Phys. World {\bf 10}, 43 (1997); Percival I.C., Strunz W.T.:
Proc. R. Soc. A {\bf 453}, 431 (1997)
\bibitem{hughston} Hughston L.: Proc. R. Soc. A {\bf 452}, 953 (1996) 
\bibitem{pearle} Pearle P.: Phys. Rev. {\bf 33}, 2240 (1986); 
Pearle P., Squires E.: Found. Physics {\bf 26}, 291 (1996) 
\bibitem{anast} Anastopoulos C.: Phys. Rev. D {\bf 54}, 1600 (1996)
\bibitem{lamine} Lamine B.,  Herv\a'e R., Lambrecht A., Reynauld S.: Phys. Rev. Lett.
{\bf 96}, 050405 (2006)
\bibitem{breuer} Breuer H.-P., G\"okl\"u  E., L\"ammerzahl C.: Class. Quant. Grav. {\bf 26}, 105012 (2009)
\bibitem{add4} Arkani-Hamed  N., Dimopoulos S., Dvali G., Kaloper N.: Phys. Rev. Lett. {\bf 84}, 586 (2000);
Arkani-Hamed N., Dimopoulos S., March-Russell J.: Phys. Rev. D {\bf 63}, 064020 (2001)
\bibitem{dice2010-more} Doyen G., Drakova D.: J. Phys.: Conf. Ser. {\bf 306}, 012033 (2011)
\bibitem{zwiebach} Zwiebach B.: A First Course in String Theory. Cambridge Univ. Press (2004)
\bibitem{misner}  Misner Ch.W., Thorne K.S., Wheeler J.A.: Gravitation. San Francisco: Freeman W.H. (1973)
\bibitem{zee} Zee A.: Quantum Field Theory in a Nutshel. Princeton Univ. Press (2003)
\bibitem{doyenDICE2012} G. Doyen and D. Drakova, J. Phys.: Conf. Ser. {\bf 442}, 012032 (2013)
\bibitem{drakovaDICE2012} Drakova  D., Doyen G.: J. Phys.: Conf. Ser. {\bf 442}, 012049 (2013)
\bibitem{arxive} Drakova D., Doyen G.: http://arxiv.org/quant-ph/pdf/1204.5606 (2012)
\bibitem{lippschwi} Lippmann B.A. Schwinger J.:  Phys. Rev. {\bf 79}, 469 (1950)
\bibitem{gradshteyn} Gradshteyn I.S., Ryzhik I.M.: Table of Integrals, Series, and Products, p. 312 (integral 3.354.5). Academic Press, New York (1965)
\bibitem{bell} Bell J.: Beables for Quantum Field Theory. In: Bell J.S.: Speakable
and Unspeakable in Quantum Mechanics, p. 171-180. Cambridge Univ. Press (1987)
\bibitem{Bell2} Bell J.: Physics {\bf 1},195 (1965); Rev. Mod. Phys. {\bf 38}, 447 (1966)
\bibitem{kant} Kant I.: Critique of Pure Reason. 1781, tr. by N. Kemp Smith. St. Martin's Press, New York (1965)
\bibitem{auyang} Auyang S.Y.: How is quantum field theory possible?, p. 71. Oxford Univ. Press (1995), see also notes 69 and 120.
\bibitem{callender} Callender C.: Finding "real" time in quantum mechanics. In: Craig W.L., Smith Q. (eds.) Einstein,
Relativity and Absolute Simultaneity, p. 50. Routledge (2011)
\bibitem{EPR} Einstein A., Rosen N., Podolsky B., Phys. Rev. {\bf 47}, 777 (1935)
\bibitem{tooley} Tooley M.: A defense of absolute simultaneity.
In: Craig W.L., Smith Q. (eds.) Einstein, Relativity and Absolute Simultaneity,
p. 229-243. Routledge, (2011)
\bibitem{albert} Albert D.Z.: Quantum Mechanics and Experience, p. 155-160. Harvard Univ. Press, (1992)
\bibitem{ryder} Ryder L.: Introduction to general relativity, p. 311. Cambridge Univ. Press (2009)
\bibitem{cheng2} Ta-Pei Cheng: Relativity, Gravitation and Cosmology - A basic Introduction. 2nd Edition,
Oxford Master Ser. Physics. Oxford Univ. Press (2010)
\bibitem{bohm} Bohm D.: A Suggested Interpretation of the Quantum Theory in Terms of
''Hidden Variables''.  Phys. Rev. {\bf 85}, 165 (1952) and Phys. Rev.
{\bf 85}, 180 (1952)
\end{thebibliography}
\end{document}